\documentclass[10pt]{article}
\usepackage{dp_delphititle}
\usepackage{lineno}
\usepackage{graphicx}
\usepackage{amssymb,amsfonts,amsmath,amsthm}
\usepackage{cite}
\usepackage{float}
\usepackage{xspace}
\usepackage{psfrag}
\usepackage{multirow}

\def\DpAuthors{ATLAS Collaboration}
\def\DpSubmit{}
\def\DpTitle{Study of ATLAS sensitivity to FCNC top decays}
\def\DpComment{Send comments to:}
\def\DpComment{}
\def\DpEMail{
jcarlos@mars.fis.uc.pt, 
nuno.castro@cern.ch,
filipe.veloso@cern.ch, 
antonio.onofre@cern.ch
}
\def\DpEMail{}

\begin{document}

\makeatletter
\input{coll.sty}
\makeatother


\begin{titlepage}
\pagenumbering{roman}

\CERNpreprint{SN-ATLAS-2007-059}{}   
\date{{\small 13 September 2007}}                      
\title{\DpTitle}                            
\address{\DpAuthors}                        


\addtolength{\textheight}{10mm}
\addtolength{\footskip}{-5mm}
\begingroup
%
\newcommand{\DpName}[2]{\hbox{#1$^{\ref{#2}}$},}
\newcommand{\DpNameTwo}[3]{\hbox{#1$^{\ref{#2},\ref{#3}}$},}
\newcommand{\DpNameThree}[4]{\hbox{#1$^{\ref{#2},\ref{#3},\ref{#4}}$},}
\newskip\Bigfill \Bigfill = 0pt plus 1000fill
\newcommand{\DpNameLast}[2]{\hbox{#1$^{\ref{#2}}$}}

\begin{center}
  \noindent
  {\large 
\DpName{J.Carvalho}{LIP}
\DpName{N.Castro}{LIP}
\DpName{L.Chikovani}{GEO}
\DpName{T.Djobava}{TBI}
\DpName{J.Dodd}{CUNL}
\DpName{S.McGrath}{CUNL} 
\DpName{A.Onofre}{LIP}
\DpName{J.Parsons}{CUNL}
\DpNameLast{F.Veloso}{LIP}

}
\end{center}

\normalsize
\endgroup
\titlefoot{LIP - Dep. F\'\i sica, Universidade de Coimbra, 3004-516 Coimbra, 
Portugal
    \label{LIP}}
\titlefoot{E.Andronikashvili Institute of Physics, Tbilisi, Georgia
    \label{GEO}}
\titlefoot{High Energy Physics Institute, Tbilisi State University, Georgia
    \label{TBI}}
\titlefoot{CU, Columbia University, Nevis Laboratories, 136 South Broadway, 
P. O. Box 137, Irvington, NY 10533, USA
    \label{CUNL}}

\addtolength{\textheight}{-10mm}
\addtolength{\footskip}{5mm}


\begin{shortabs}                            
\noindent

The ATLAS experiment sensitivity to top quark Flavour Changing Neutral Current (FCNC) decays 
was studied at LHC using $t\bar t$ events.  
While one of the top quarks is expected to follow the
dominant Standard Model decay $t \to b W$, the other decays through a FCNC
channel, \emph{i.e.} $t\to Z u(c)$, $t\to \gamma u(c)$ or $t\to g u(c)$.
Different types of analyses, applied to each FCNC decay mode, were compared. The FCNC branching ratio
sensitivity (assuming a $5\sigma$ signal significance) and 95\% confidence level limits on the branching
ratios (in the hypothesis of signal absence) were obtained.

\end{shortabs}

\vfill

\begin{center}
\DpSubmit \ \\          
\DpComment \ \\
\DpEMail{filipe.veloso@lipc.fis.uc.pt} \ \\
\end{center}

\vfill
\clearpage

\headsep 10.0pt

\headsep 30.0pt
\end{titlepage}

  \pagenumbering{arabic}
  \setcounter{footnote}{0}
  \large
\section{Introduction \label{introduction}}

Flavour Changing Neutral Currents (FCNC) are strongly suppressed in the
Standard Model (SM) due to the Glashow-Iliopoulos-Maiani (GIM)
mechanism~\cite{GIM}. Although absent at tree level, small FCNC contributions
are expected at one loop level, according to the Cabibbo-Kobayashi-Maskawa
(CKM) mixing matrix~\cite{SMFCNC}. In the top quark sector of the SM, these
contributions limit the FCNC decay branching ratios to the gauge bosons,
$BR_{t\to qX}$ ($X=Z,\gamma,g$), to below $10^{-10}$.  There are however
extensions of the SM, like supersymmetry (SUSY)~\cite{supers}, multi-Higgs
doublet models \cite{MHiggs} and SM extensions with exotic (vector-like)
quarks~\cite{exotics}, which predict the presence of FCNC contributions
already at tree level and significantly enhance the FCNC decay branching
ratios compared to the SM predictions.

Due to its large mass, much higher than any other known fermion, the top
quark is a very good laboratory to look for physics beyond the SM. If the top
quark has FCNC anomalous couplings to the gauge bosons, its decay properties
would be affected, and possibly measured at colliders, in addition to the
dominant decay mode $t \to b W$. Indeed one of the most prominent signatures
of FCNC processes at the Large Hadron Collider (LHC), would be the direct
observation of a top quark decaying into a charm or an up quark together with
a $\gamma, g$ or $Z$ boson~\cite{Aguilar-Saavedra:2004wm}. In the effective
Lagrangian approach~\cite{Buchmuller:1985jz,top39} the new top quark decay
rates to the gauge bosons~\cite{yr},
\begin{equation}
  \Gamma({ t \to q g}) = \left( \frac{\kappa^g_{tq}}{\Lambda} \right)^2
  \frac{8}{3} \alpha_s m_t^3,
\end{equation} 
\begin{equation}
  \Gamma({  t \to q \gamma})  =
  \left( \frac{\kappa_{tq}^{\gamma}}{\Lambda}\right)^2 2 \alpha  m_t^3,
\end{equation}
\begin{equation}
\Gamma({ t \to q Z})_{\gamma} = \left(|v^Z_{tq}|^2+|a^Z_{tq}|^2\right)
 \alpha m^3_t
 \frac{1}{4 M^2_Z \sin^2 2\theta_W }
 \left ( 1 - \frac{m_Z^2}{m_t^2} \right )^2
 \left ( 1 + 2\frac{m_Z^2}{m_t^2} \right )\mathrm{\ and}
\end{equation}
\begin{equation}
  \Gamma({ t \to q Z})_{\sigma} = \left( \frac{ \kappa^Z_{tq}}{\Lambda}
  \right)^2
  \alpha \, m^3_t \frac{1}{ \sin^2 2\theta_W }
  \left ( 1 - \frac{m_Z^2}{m_t^2} \right )^2
  \left ( 2 + \frac{m_Z^2}{m_t^2} \right ),
\label{anomeq:br}
\end{equation} 
can be expressed in terms of the $\kappa^g_{tq}$,
$\kappa_{tq}^{\gamma}$, $(|v^Z_{tq}|^2+|a^Z_{tq}|^2)$ and
$\kappa^Z_{tq}$ anomalous couplings to the $g$, $\gamma$ and $Z$ bosons
respectively.  The energy scale associated with this new physics is
represented by $\Lambda$, while $\alpha_s$ and $\alpha$ are,
respectively, the strong and electromagnetic coupling constants. The
electroweak mixing angle is represented by $\theta_W$ and the top and
$Z$ masses are represented, respectively, by $m_t$ and $ m_Z$.

Although FCNC processes associated with the production \cite{LEP,HERA} and
decay\cite{CDF} of top quarks have been studied at colliders ($BR_{t\to Z
q}<33\% $ and $BR_{t\to \gamma q}<3.2\%$ at 95\%~Confidence
Level~(CL)\cite{CDF}), the amount of top quark relevant data collected up to
now is not comparable with the statistics expected at the LHC.  The LHC will
operate with a centre-of-mass energy of 14~TeV, and in the low luminosity
phase ($l=10^{33}$~cm$^{-2}$s$^{-1}$), several millions of top quarks will be
produced per year and experiment, mainly in pairs (with a NLO cross-section
of $833$~pb~\cite{yr,ttcs}), but also through single top production (with an
expected NLO cross-section of 280~pb~\cite{tcs}).

\begin{figure}
  \begin{center}
    \includegraphics[width=\textwidth]{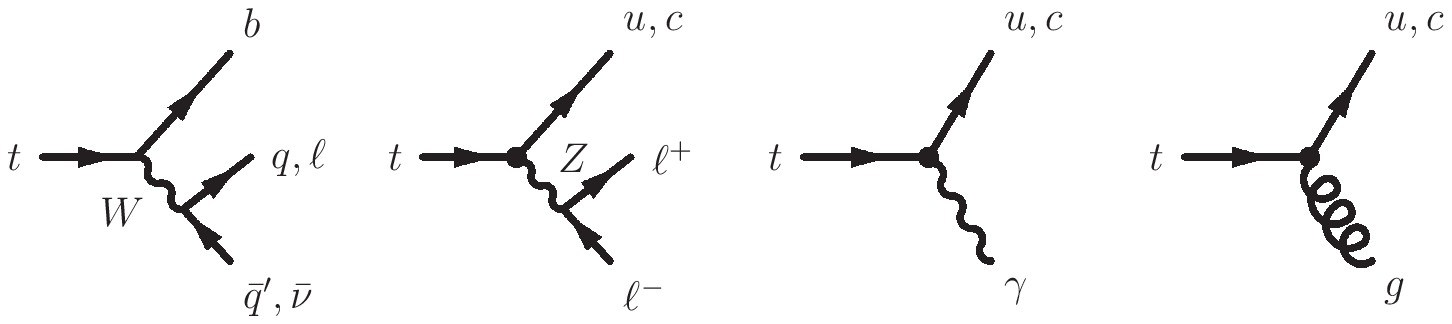}
    a)\hspace{.23\textwidth} b)\hspace{.23\textwidth}
    c)\hspace{.23\textwidth} d)
    \vspace{.5em}
  \end{center}

  \vspace*{-2em}
  \emph{\caption{Feynman diagrams for the top quark decays considered in this 
  paper:
  a)~SM decay $t\to bW$;
  b)~FCNC decay $t\to Zq$;
  c)~FCNC decay $t\to\gamma q $ and
  d)~FCNC decay $t\to gq$.
  }
  \label{fig:feyndec}}
\end{figure}

This paper is devoted to the study of the ATLAS experiment~\cite{ATLAS}
sensitivity to FCNC top quark decays at the LHC. While one of the top
quarks is expected to follow the dominant SM decay ($t \to b W$), the
other decays through a FCNC channel, \emph{i.e.} $t\to Z q$, $t\to
\gamma q$ or $t\to g q$. The corresponding Feynman diagrams 
are shown in Fig.~\ref{fig:feyndec}. Different types of analyses (cut-based
and likelihood-based) were applied to each FCNC decay mode and their results
compared.

This paper is organised as follows. After the introduction, a
description of the simulated signal and background is given in
section~\ref{simulation}. The analysis criteria applied to each FCNC
channel are described in section~\ref{analyses} and in
section~\ref{results} a comparison of the results obtained by the
different analyses~\cite{chikovani,carvalho} is presented within two
different approaches: branching ratio sensitivities (assuming a
$5\sigma$ signal significance for discovery) and 95\% confidence level
limits (in the hypothesis of signal absence). These results are compared
with previously published ones~\cite{LEP,HERA,CDF,ATLAS,dodd,cakir}. In 
section~\ref{conclusions} the conclusions are presented.

\section{Signal and background simulation \label{simulation}}

The Monte Carlo (MC) generation of the QCD ($b\bar{b}$), $W+$jets,
$Z/\gamma^*+$jets, $WW$, $ZZ$ and $ZW$ background processes was done with the
library PYTHIA~\cite{pythia}. Single top quark production was generated with
TopReX 4.05~\cite{toprex}, and the SM top pair production ($t\bar{t}_{SM}$)
was generated using TopReX and PYTHIA. These libraries were also used to
generate signal $t\bar t$ samples, where one of the $t$-quarks decays via
Charged Currents (CC) into $bW$ and the other one decays through FCNC into
$qZ$, $q\gamma$ or $qg$. For TopReX the anomalous couplings to the $g$,
$\gamma$ and $Z$ bosons were set to $\kappa^g_{tq}=\kappa_{tq}^{\gamma}=
(|v^Z_{tq}|^2+|a^Z_{tq}|^2)^{1/2}=\kappa^Z_{tq}=0.1$ and $\Lambda$ was set to
1~TeV. The top mass was set to $175$~GeV$/c^2$. Different values,
$170$~GeV$/c^2$ and $180$~GeV$/c^2$, were also considered for the study of
systematic uncertainties, as explained in section~\ref{systematics}.
No SUSY backgrounds or other contributions beyond the SM were considered in 
the present analyses. 
The CTEQ2L and CTEQ5L Parton Distribution Functions (PDF)
were used~\cite{pythia,toprex} in the analyses and the CTEQ4M was used for 
systematic studies. No pile-up was taken into account.


The generated background and signal events were passed through the ATLAS fast
simulation packages ATLFAST~\cite{atlfast} and ATLFASTB~\cite{atlfast}. For
each event, these packages begin by simulating the energy deposition in the
calorimeter cells of all the stable particles.  The calorimeter cells are
clustered within a cone of $\Delta R =
\sqrt{(\Delta\phi)^2 + (\Delta\eta)^2} = 0.4$. Cells with $E_T>1.5$~GeV are
used as cluster seeds and the cone algorithm is applied in decreasing order
of $E_T$. Only clusters with $E_T>5$~GeV are considered. The polar angle and
the momentum of photons are smeared according to Gaussian parameterizations.
For electrons, their momenta are smeared according to a Gaussian
parameterizations. The momentum of each muon is smeared according to a
resolution which depends on the $p_T$, $|\eta|$ and $\phi$. The photon
(electron) energy resolution is $\delta E/E < 2.9\%$~ ($3.3\%$), for
$E>20$~GeV. The transverse momentum resolution of muons with
$p_T<100$~GeV$/c$ is $\delta p_T/p_T\lesssim 2\%$. Photons, electrons and
muons are selected only if they have $|\eta|<2.5$ and $p_T>5$~GeV$/c$
($p_T>6$~GeV$/c$ for muons). They are classified as isolated if the 
transverse
energy of the cluster associated to the particle, inside a cone of $\Delta R
= 0.2$, does not exceed $10$~GeV the particle energy, and the $\Delta R$ 
from 
other energy clusters must be above 0.4. The clusters of 
energy
depositions not associated to isolated photon, electrons or muons are used
for the jet reconstruction. Their momenta are smeared according to a Gaussian
distribution which depends on $|\eta|$. Jets are selected if they have
$E_T>10$~GeV. For $E>20$~GeV, the jet energy resolution is less than 12\%
($|\eta| < 3$) and less than 24\% ($|\eta| > 3$). The missing transverse
momentum is estimated by summing the transverse momentum of the isolated
photons, electrons, muons and jets. The non isolated muons and the clusters
of energy deposition not associated to isolated photons, electrons, muons or
jets, are also taken into account. In the ATLAS detector, it will be possible
to identify $b$-jets with $|\eta|<2.5$ by using $b$-tagging tools. The
algorithm was simulated by setting a $b$-tagging efficiency to 60\%, with
contamination factors set to 14.9\% (1.1\%) for $c$-jets (light quark, gluon
and tau jets) for the low luminosity phase ($l=10^{33}$~cm$^{-2}$s$^{-1}$). 
In order to check the dependence of the analysis with the $b$-tagging
efficiencies, different values, 50\% and 70\% (corresponding to the expected
$b$-tag variation within the interesting signal transverse momentum range),
were also considered for the systematic studies and the high luminosity phase
($l=10^{34}$~cm$^{-2}$s$^{-1}$), with contamination factors of 9.2\% (0.4\%)
and 23.3\% (2.9\%) for $c$-jets (light quark, gluon and tau jets),
respectively.

Initial and final state QED and QCD radiation (ISR+FSR), multiple
interactions and hadronization were taken into account in the event
generation. Due to the hadronization and FSR, the jets are reconstructed with
less energies than those from the original quarks or gluons. The jets
energies were calibrated by the ATLFASTB package, by applying a calibration
factor, $K^{\mathrm{jet}}=p_T^{\mathrm{parton}}/p_T^{\mathrm{jet}}$, that is
the ratio between the true parton energy and the reconstructed jet energy.
The calibration factor depends on the $p_T$ and is different for $b$-tagged
and light jets.

Preliminary full simulation studies, based on the ATHENA
framework\cite{athena}, indicate a fair agreement between the fast and full
simulations of the ATLAS detector.

\section{Topologies and event selection \label{analyses}}

The $t\bar t$ final states corresponding to the different FCNC top decay
modes lead to different topologies according to the number of jets, leptons
and photons. There is however a common characteristic of all channels under
study, \emph{i.e.} in all of them one of the top quarks is assumed to decay
through the dominant SM decay mode $t\to bW$ and the other is forced to decay
via one of the FCNC modes $t\to Z q$, $t\to \gamma q$ or $t\to g q$. Two
different types of analyses, labelled ``cut-based'' and ``likelihood-based'',
were used to study the ATLAS sensitivity to FCNC top quark decays.  For both
analyses the leptonic decays of the $W$ ($W \to \ell \nu_\ell$) were taken
into account\footnote{For the cut-based analyses $\ell=e,\mu$, while for the
likelihood-based analyses $\ell=e,\mu,\tau$}. In addition, for the FCNC
channel $t\to Z q$, the hadronic decay of the $W$ ($W \to qq'$) was also
considered for the cut-based analysis.

\subsection{$\boldsymbol{t\to Z q}$ channel}

The QCD backgrounds at hadron colliders make the search for the signal via
the fully hadronic channel (when both the $W$ and $Z$ decay hadronically)
very difficult. For this reason only the leptonic decay of the $Z$ was
considered. The final state was then determined by the decays of the $W$
boson. Two different possible decay channels have been considered: the first
('leptonic mode') where the $W$ decays leptonically $W\to\ell\nu$, and the
second ('hadronic mode') with $W\to jj$. The hadronic $W$ decay signature has
a larger branching fraction, but suffers from larger backgrounds. The
experimental signature of the leptonic mode includes three isolated charged
leptons, two of which reconstruct a $Z$ boson, large missing transverse
energy due to the neutrino and at least two jets, one of which is tagged as a
$b$-jet. The signature of the hadronic mode is characterised by having two
leptons (again with $m_{\ell^{+}\ell^{-}}\approx m_Z$) and at least four
jets, one of which is tagged as a $b$-jet. Following a previous
analysis~\cite{dodd}, new cut-based~\cite{chikovani} and
likelihood-based~\cite{carvalho} analyses were developed and are described
below.

\subsubsection{Cut-based analysis: hadronic mode}

The final state for the hadronic $W$ decay mode is $t\bar{t} \rightarrow
Zq Wb \rightarrow \ell^+\ell^-j jjb$. This mode has the following
backgrounds: $Z+$jets production, followed by the decay $Z \rightarrow
\ell^+\ell^-$, $pp \rightarrow W^{\pm}Z+X \rightarrow jj \ell^+\ell^-
+X$, and $t\bar{t} \rightarrow WbWb$ with the final state topologies (a)
$\ell^+\nu b \ell^- \bar {\nu}b$, or (b) $\ell^{\pm} \nu b jjb$. In the
case of (a), the additional two jets must come from QCD radiation, while
in (b) the source of leptons is from cascade decays. $Z+$jets production
at the LHC has a relatively large cross-section, dominated by $qg
\rightarrow Zq$ and $q\bar{q} \rightarrow Zg$ processes. To decrease the
size of the background sample which needed to be generated, thresholds were
imposed at the generator level on the invariant mass,
${m}=\sqrt{\hat{s}}>130$~GeV$/c^2$, where $\sqrt{\hat{s}}$ is the effective
centre-of-mass energy, and transverse momentum, ${p_T}>50$~GeV$/c$, of the
hard scattering process. The cross-section for this subsample of events was
$\sigma_{Z+jets}$=3186~pb.  The $WZ$ background is the electroweak process
$pp \rightarrow W^{\pm}Z+X$, and has an assumed cross-section of $\sigma_{WZ}
= 26.58$~pb. Background samples of $2.1\times 10^7$ $Z+$jets events,
$1.2\times 10^5$ $WZ$ events and $2.8\times 10^7$ $t\bar{t}$ events were
generated. Assuming the above mentioned production cross-sections, and
including the relevant branching ratios, these background samples correspond
to an integrated luminosity of 100~fb$^{-1}$.

The analysis began with preselection cuts requiring that the event contains
at least two charged leptons (electrons with $p_T>5$~GeV$/c$ within
pseudorapidity $|\eta|<2.5$ and muons with $p_T>6$~GeV$/c$ within
pseudorapidity $|\eta|<2.4$), and include a pair of opposite-sign and
same-flavour leptons, compatible with them having come from a $Z$ decay. In
addition, the number of jets with ${p_T}_{\mathrm{jet}} > 15$~GeV$/c$ within
pseudorapidity $|\eta|<5.0$ was required to be at least four.
 After preselection cuts, 46\% of the signal events were accepted, 
while only 3.0\%, 3.5\% and 4.1\% of the $t\bar{t}$, $Z+$jets and $WZ$ 
background events, respectively, were selected.

The next cuts required the presence of two isolated leptons with 
${p_T}_{\ell}>20$~GeV$/c$ and the demand for at least four jets with 
${p_T}_{\mathrm{jet}}>50$~GeV$/c$ and $|\eta^{j}|<2.5$. The isolation $\Delta 
R$ of the jets (measured in relation to other jets and leptons) was then 
required to be greater than 0.4.

\begin{figure}
  \begin{center}
    \includegraphics[width=.995\textwidth]{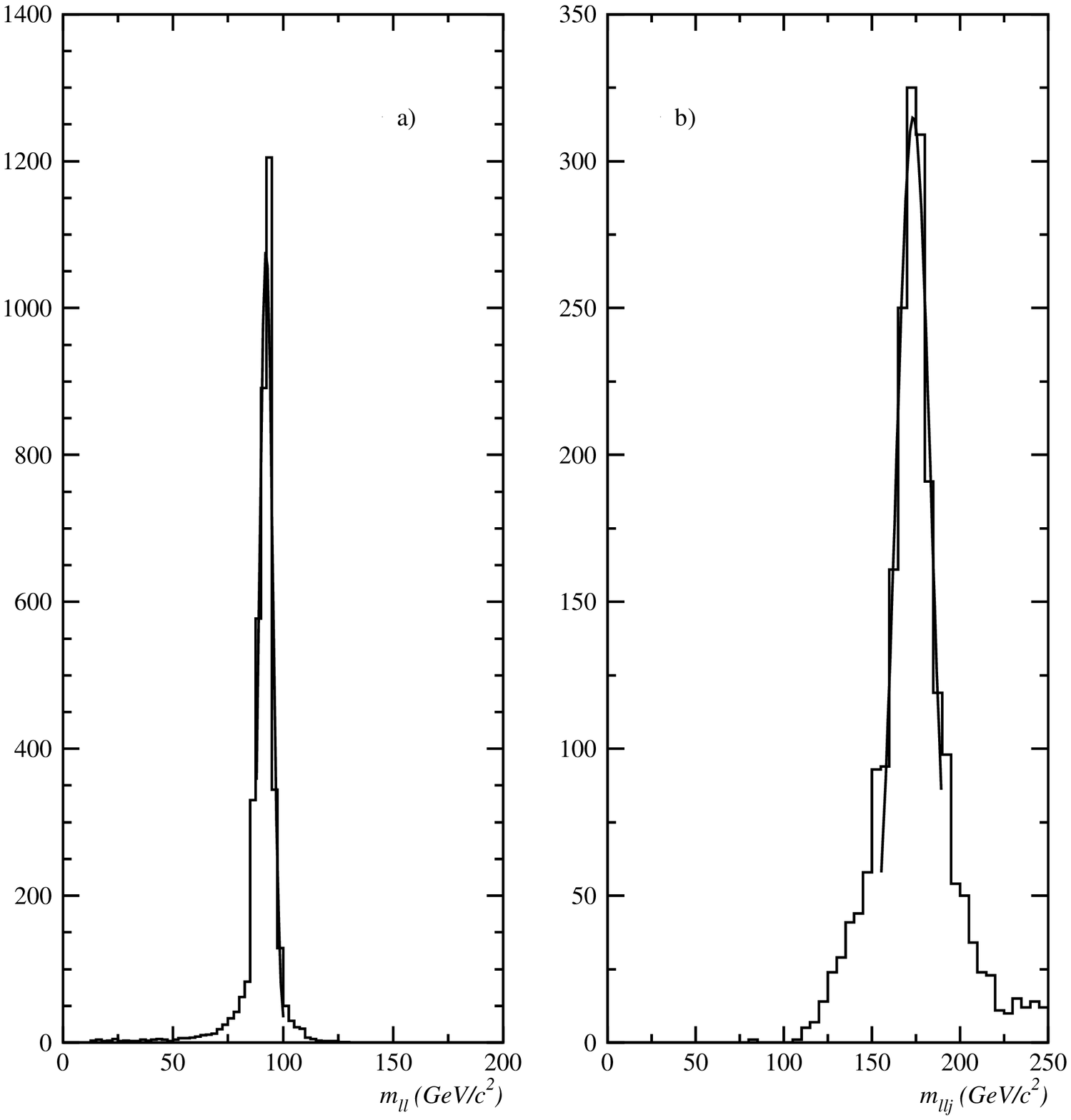}
  \end{center}
  \vspace{-1em}

  \hspace{.24\textwidth}a)\hfill b)\hspace{.24\textwidth}\vspace{.5em}

  \vspace*{-2em}
  \emph{\caption{Distributions for the hadronic mode of a) 
  reconstructed invariant mass of the 
  lepton pairs, $m_{\ell\ell}$ for the best combination and b) 
  reconstructed invariant mass of $t \rightarrow \ell\ell j$ 
  for the best combination of $llj$.}
  \label{fig:mljhad}}
\end{figure}

Fig.~\ref{fig:mljhad} presents the distributions of reconstructed dilepton
invariant mass and of reconstructed $t\rightarrow Zq$ invariant mass
$m_{\ell\ell j}$ for the best combinations of $\ell\ell j$ (i.e. $\ell\ell j$
combination with the closest to top mass value of invariant mass $m_{\ell\ell
j}$) for the signal sample. A cut was then applied on the dilepton invariant
mass,requiring that it lie within $\pm$6~GeV$/c^2$ around $m_Z$. 

\begin{figure}
  \begin{center}
    \includegraphics[width=.995\textwidth]{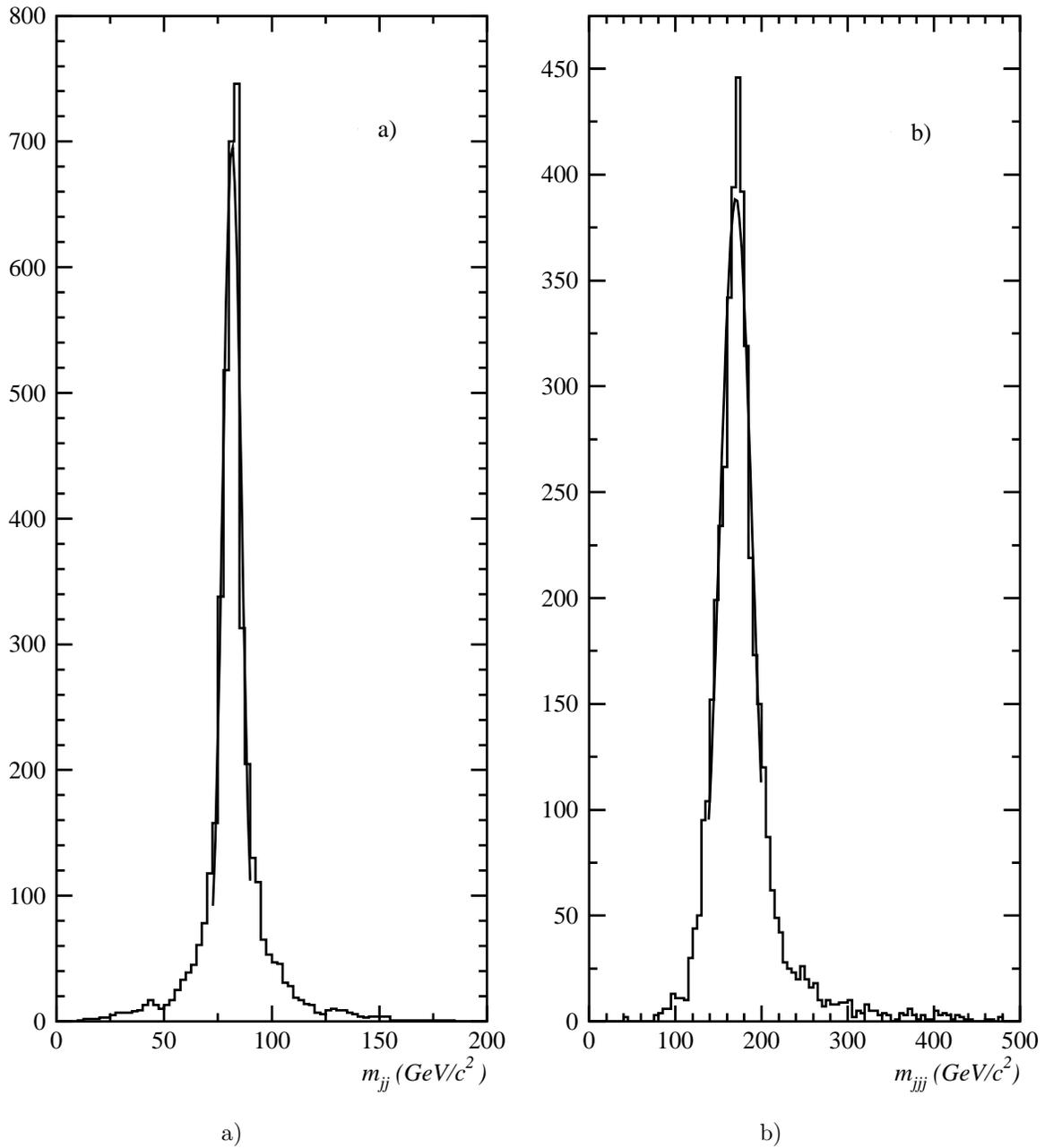}
  \end{center}
  \vspace{-1em}

  \hspace{.24\textwidth}a)\hfill b)\hspace{.24\textwidth}\vspace{.5em}

  \vspace*{-2em}
  \emph{\caption{Distributions for the hadronic mode of a) reconstructed 
  invariant mass of the jet pairs, $m_{jj}$ for the best combination and 
  b) reconstructed invariant mass of $t \rightarrow jjb$ 
  for the best combination of $jjb$.}
  \label{fig:mjjhad}}
\end{figure}

To suppress the large remaining $Z+$jets background, it was necessary to use
the information that signal events contain, in addition to the decay
$t\rightarrow Zq$, a hadronic decay $t\rightarrow Wb\rightarrow jjb$ of the
other top quark. The hadronic top quark decay was, therefore, reconstructed
as part of the signal requirement. First, a pair of jets was required to have
an invariant mass $m_{jj}$ within a 16~GeV$/c^2$ around $m_W$.
Fig.~\ref{fig:mjjhad} shows the distribution of reconstructed $m_{jj}$ for
the best combinations of $jj$ (i.e. $jj$ combination with the closest to $W$
mass value of invariant mass $m_{jj}$) for the signal events. The $W$ mass
resolution is $\sigma_{m_{jj}}=8$~GeV$/c^2$. A requirement was then made to
have exactly one jet tagged as a $b$-jet. Finally,the $jjb$ invariant mass
was required to lie within 8~GeV$/c^2$ around $m_t$.

Fig.~\ref{fig:mjjhad} presents the distribution of the reconstructed
invariant top mass ($m_{jjb})$ for the best combinations of $jjb$ (i.e. $jjb$ 
combination with the closest to top mass value of invariant mass 
$m_{jjb}$) for the signal. The top mass resolution is $\sigma
(m_{jjb})=18.5$~GeV$/c^2$, implying that the mass window applied is
rather narrow in order to get a large background rejection. The sequence
of cuts required to reconstruct the hadronic decay of the other top
quark dramatically suppresses the backgrounds, but also reduces the
signal efficiency by almost an order of magnitude. For those events with
an accepted $t\rightarrow jjb$ candidate, the invariant mass of the $Z$
candidate with the remaining unassigned high $p_T$ jets was
reconstructed to look for a signal from $t\rightarrow Zq$ decays. The
resolution $\sigma$ of $m_{\ell\ell j}$ distribution is $\sigma_{m_{\ell\ell
j}}=9.9$~GeV$/c^2$ (see Fig.~\ref{fig:mljhad}). The analysis cuts reduce
the $WZ$ background to a negligible level in the $m_{Zq}\pm
24$~GeV$/c^2$ mass window. Two events of the $Z+$jets background are
accepted in this mass window.

Table~\ref{tab:hadronic} summarises the effects of the sequential
application of the above described various analysis cuts
on the background samples and on the sample of 19000
signal events of the topology $t\bar{t} \rightarrow ZqWb \rightarrow
\ell^+\ell^-j jjb$.

\begin{table}[bt]
\begin{center}
  \begin {tabular}{|l||r||r|r|r|r|}
    \hline
    \multicolumn{1}{|c||}{Description} & \multicolumn{1}{|c||}{Signal}             & \multicolumn{4}{|c|}{Background Processes}\\
    \cline{3-6}
    \multicolumn{1}{|c||}{of}          & \multicolumn{1}{|c||}{$t\rightarrow Zq$}  & \multicolumn{1}{|c|}{$Z+$jets} & \multicolumn{1}{|c|}{$Z+W$} & \multicolumn{1}{|c|}{$t\bar t$} & \multicolumn{1}{|c|}{$t\bar t$}\\
  \multicolumn{1}{|c||}{Cuts}    & &  &  &di-leptonic & {semi-leptonic} \\
    \cline{2-6}
    \multicolumn{1}{|c||}{   }        & \multicolumn{1}{|c||}{$\varepsilon$ (\%)} & \multicolumn{1}{|c|}{Nevt} & \multicolumn{1}{|c|}{Nevt} & \multicolumn{1}{|c|}{Nevt} & \multicolumn{1}{|c|}{Nevt} \\
    \hline
    \hline

    Preselection                         & 46.0 & 7.5$\times 10^{5}$ &               4970 &  5.8$\times 10^{5}$ &  2.7$\times 10^{5}$ \\
    2 leptons, 4 jets                    &      &                    &                    &                     &                     \\
    \hline
    2 leptons, $p_{T}^{\ell}>20$~GeV$/c$ & 37.7 & 5.9$\times 10^{5}$ &               4456 &              428800 &               11200 \\
    \hline
    4 jets, $P_{T}^{jet}>50$~GeV$/c$     & 15.2 &              63478 &                400 &               35530 &                 870 \\
    \hline
    ${\Delta}R_{jj} > 0.4 $              & 14.9 &              60421 &                390 &               35370 &                 830 \\
    \hline
    ${\Delta}R_{lj} > 0.4 $              & 14.9 &              60394 &                361 &               35370 &                 830 \\
    \hline
    $m_{Z} \pm 6$~GeV                    & 12.8 &              50973 &                268 &                3104 &                  60 \\
    \hline
    $m_{W} \pm 16$~GeV                   &  5.3 &              14170 &                139 &                 719 &                  37 \\
    \hline
    one $b$-tag                          &  2.2 &               1379 &                 11 &                 376 &                  15 \\
    \hline
    $m_{Wb} = m_t \pm 8$~GeV             &  0.6 &                 90 &                  1 &                  28 &                   4 \\
    \hline
    $m_{Zq} = m_t \pm 24$~GeV            &  0.4 &                  2 &                  0 &                   5 &                   0 \\
    \hline
  \end{tabular}
  \emph{
  \caption{The number of events, normalised to $L=100$~fb$^{-1}$, and
  efficiency ($\%$) of selection cuts applied in sequence for the signal
  and backgrounds in the hadronic decay mode in the $t\to Zq$ channel,
  obtained with a cut-based analysis, are shown.
  }
  \label{tab:hadronic}}
\end{center}
\end{table}

\subsubsection{Cut-based analysis: leptonic mode}

The final state for the leptonic decay mode is $t\bar{t} \rightarrow
ZqWb \rightarrow\ell^+\ell^-j \ell\nu b$. The experimental signature
therefore includes three isolated charged leptons, two of which
reconstruct a $Z$ boson, and large missing transverse energy due to the
undetected neutrino.

This mode has the following backgrounds: $Z(\to\ell\ell)+$jets, $pp
\rightarrow W^{\pm}Z+X \rightarrow \ell^{\pm}\nu\ell^+\ell^- +X$, and
$t\bar{t} \rightarrow W^{+}bW^{-} \bar{b} \rightarrow \ell^{+} \nu b
\ell^{-} \bar{\nu} b$. Assuming the production cross-sections given
earlier, and including the relevant branching ratios, background samples
of $2.1\times 10^7$~$Z+$jets events, 38000~$WZ$ events, and $3.9\times
10^6$~$t\bar t$ events were generated. These background samples
correspond to an integrated luminosity of 100~fb$^{-1}$.

Preselection cuts were first applied, requiring the presence of at least
three charged leptons (electrons with $p_T > 5$~GeV$/c$ and muons with
$p_T > 6$~GeV$/c$) within pseudorapidity $|\eta|<2.5$. Of these, at
least one pair of leptons must be of opposite sign and same flavour,
compatible with them being produced from a $Z$ decay.  In addition, the
number of jets in the event with ${p_T}_{\mathrm{jet}}>15$~GeV$/c$ within 
pseudorapidity $|\eta|<5.0$ was required to be at least two. The 
requirement of three leptons reduces significantly the $Z+$jets and 
$t\bar{t}$ backgrounds, while the requirement of two jets reduces 
significantly $WZ$ and $Z+$jets backgrounds.

The lepton criteria were then tightened, by requiring the presence of at
least three isolated, charged leptons (electrons or muons) with
${p_T}_{\ell}>20$~GeV$/c$.  The next requirement, namely that the missing
transverse momentum in the event satisfies $\not\!p_T>30$~GeV$/c$, is
effective at further reducing the $Z+$ jets background while having little
impact on the signal and other background sources. Next, it was demanded that
there be at least two jets with ${p_T}_{\mathrm{jet}} > 50$~GeV$/c$,
$|\eta_{\mathrm{jet}}|<2.5$, and satisfying the following isolation
conditions: ${\Delta}R_{jj} > 0.4$ (jet-jet isolation) and ${\Delta}R_{\ell
j} > 0.4$ (lepton-jet isolation). The cut requiring the presence of two or
more jets in each event effectively suppresses the $WZ$ background.

\begin{figure}
  \begin{center}
    \includegraphics[width=.995\textwidth]{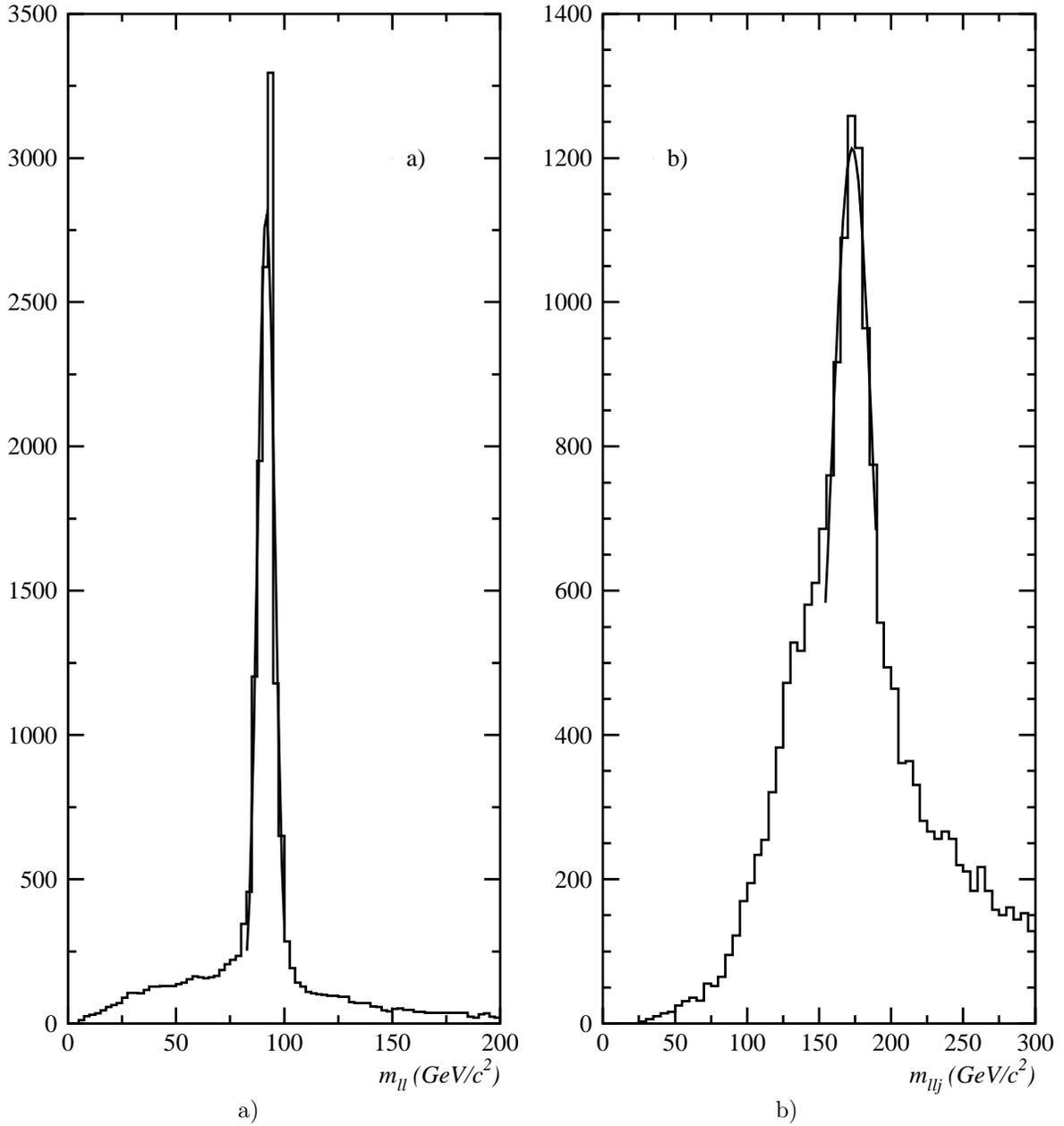}
  \end{center}
  \vspace{-1em}

  \hspace{.24\textwidth}a)\hfill b)\hspace{.24\textwidth}\vspace{.5em}

  \vspace*{-2em}
  \emph{\caption{Distributions for the leptonic mode of a) 
  reconstructed invariant mass of the lepton pairs, $m_{\ell\ell}$ and b) 
  reconstructed invariant mass of $t \rightarrow Zq \rightarrow \ell\ell j$.}
  \label{fig:mljlep}}
\end{figure}

The presence of a reconstructed $Z \rightarrow \ell^+\ell^-$ decay is a
powerful cut against the $t \bar t$ background. A like-sign, same-flavor pair
of isolated leptons was required to reconstruct to the $Z$ mass within $m_{Z}
\pm 6$~GeV$/c^2$. Fig.~\ref{fig:mljlep} presents the distribution of
reconstructed invariant mass of $\ell\ell$ pairs $m_{\ell\ell}$, for all
dilepton combinations for the signal events. The width of the accepted
window corresponds to approximately twice the $Z$ mass resolution of about
2.9~GeV$/c^2$. The next requirement was the presence in the event of exactly
one tagged $b$-jet, which is effective at further reducing the $WZ$
background Finally, a peak at the top quark mass in the $Zj$ invariant mass
distribution was sought. In Fig.~\ref{fig:mljlep}, the distribution of
reconstructed invariant mass $m_{\ell\ell j}$ for all combinations of
$\ell\ell j$ is presented for the signal events. The top quark mass
resolution is $\sigma(m_{\ell\ell j}) = 14$~GeV$/c^2$. Accepted combinations
were required to lie within $\pm 24$~GeV$/c^2$ ($\sim ~2\sigma$) around the
known top quark mass. This top mass window removes almost completely the
remaining background and the final signal efficiency is 6.1\% with a total
background of 7 events for an integrated luminosity of 100~fb$^{-1}$.

Table~\ref{tab:leptonic} summarises the effects of the sequential
application of the above described various analysis cuts on the 
background samples and on the sample of 20565~signal events of the 
topology $t\bar{t} \rightarrow ZqWb \rightarrow \ell^+\ell^-j \ell\nu b$.

\begin{table}[tb]
\begin{center}
  \begin {tabular}{|l||r||r|r|r|}
    \hline
    \multicolumn{1}{|c||}{Description} & \multicolumn{1}{|c||}{Signal}             & \multicolumn{3}{|c|}{Background Processes}\\
    \cline{3-5}
    \multicolumn{1}{|c||}{of}          & \multicolumn{1}{|c||}{$t\rightarrow Zq$}  & \multicolumn{1}{|c|}{$Z+$jets} & \multicolumn{1}{|c|}{$Z+W$} & \multicolumn{1}{|c|}{$t\bar t$} \\
    \cline{2-5}
    \multicolumn{1}{|c||}{Cuts}        & \multicolumn{1}{|c||}{$\varepsilon$ (\%)} & \multicolumn{1}{|c|}{Nevt} & \multicolumn{1}{|c|}{Nevt} & \multicolumn{1}{|c|}{Nevt} \\
    \hline
    \hline
    Preselection         & 80.2 & 3.7$\times 10^{5}$ &  2941 & 11.7$\times 10^{5}$ \\
3 leptons, 2 jets       &  &  & &  \\
    \hline
    3 leptons, $p_{T}^{\ell}>20$~GeV$/c$            & 43.3 &                945 &  1778 &               1858 \\
    \hline
    $\not\!p_{T}>30$~GeV & 32.7 &                 80 &  1252 &               1600 \\
    \hline
    2 jets, $P_{T}^{jet}>50$~GeV$/c$ & 19.8 &                 31 &   225 &                596 \\
    \hline
    $m_{Z} \pm 6$~GeV    & 16.8 &                 24 &   180 &                 29 \\
    \hline
    one $b$-tag          &  8.2 &                 10 &    28 &                  10 \\
    \hline
    $m_t \pm 24$~GeV     &  6.1 &                  0 &     2 &                  5 \\
    \hline
  \end{tabular}
  \emph{\caption{The number of events, normalised to $L=100$~fb$^{-1}$,
  and efficiency ($\%$) of selection cuts applied in sequence for the
  signal and backgrounds for the leptonic mode in the $t\to Zq$ channel,
  obtained with a cut-based analysis, are shown.
  }
  \label{tab:leptonic}}
\end{center}
\end{table}
 
\subsubsection{Likelihood-based analysis: leptonic mode\label{sec:qz}}

The leptonic decay mode was also studied with a likelihood-based analysis. A
general selection criteria was defined for the likelihood-based analyses:
events were required to have at least one isolated lepton with
$p_T>25$~GeV$/c$ and at least two jets with $p_T>20$~GeV$/c$ in the final
state. Both the lepton and the jets were required to have $|\eta|<2.5$.
Additionally, the transverse missing momentum had to be greater than
20~GeV$/c$. Table~\ref{tab:cuts} summarises the cuts performed in the
likelihood-based analysis.

\begin{figure}
  \psfrag{0.5}{\small 1}
  \psfrag{-0.5}{\small 0}
  \psfrag{1.5}{\small 2}
  \psfrag{2.5}{\small 3}
  \psfrag{3.5}{\small 4}
  \psfrag{4.5}{\small 5}
  \psfrag{5.5}{\small 6}
  \psfrag{6.5}{\small 7}
  \psfrag{7.5}{\small 8}
  \psfrag{8.5}{\small 9}
  \includegraphics[width=.48\textwidth]{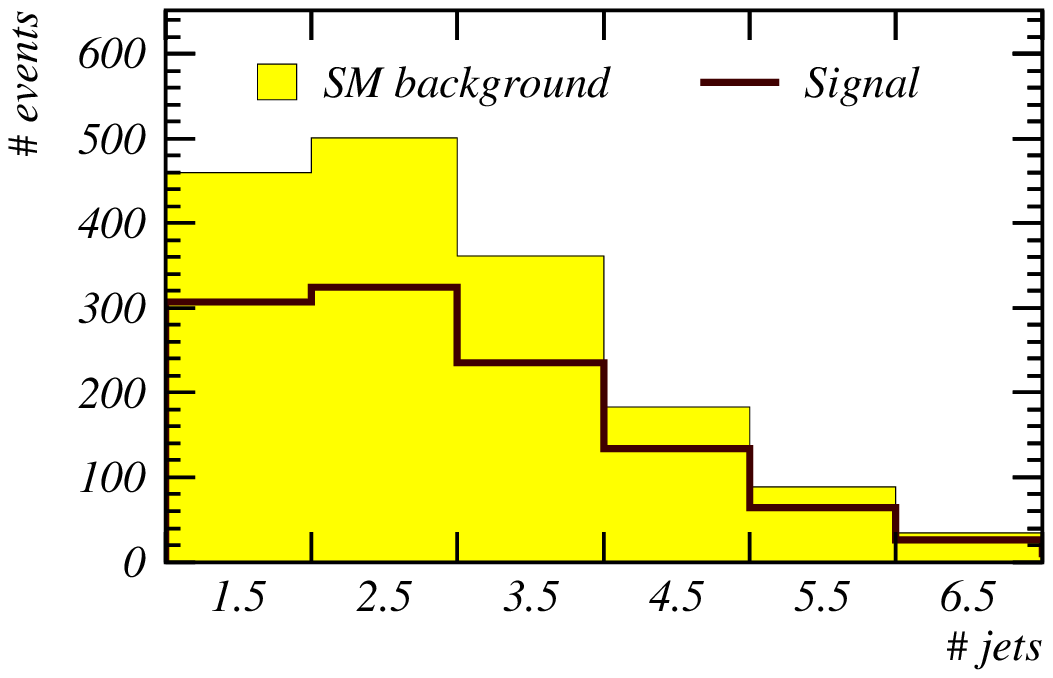}
  \hfill
  \includegraphics[width=.48\textwidth]{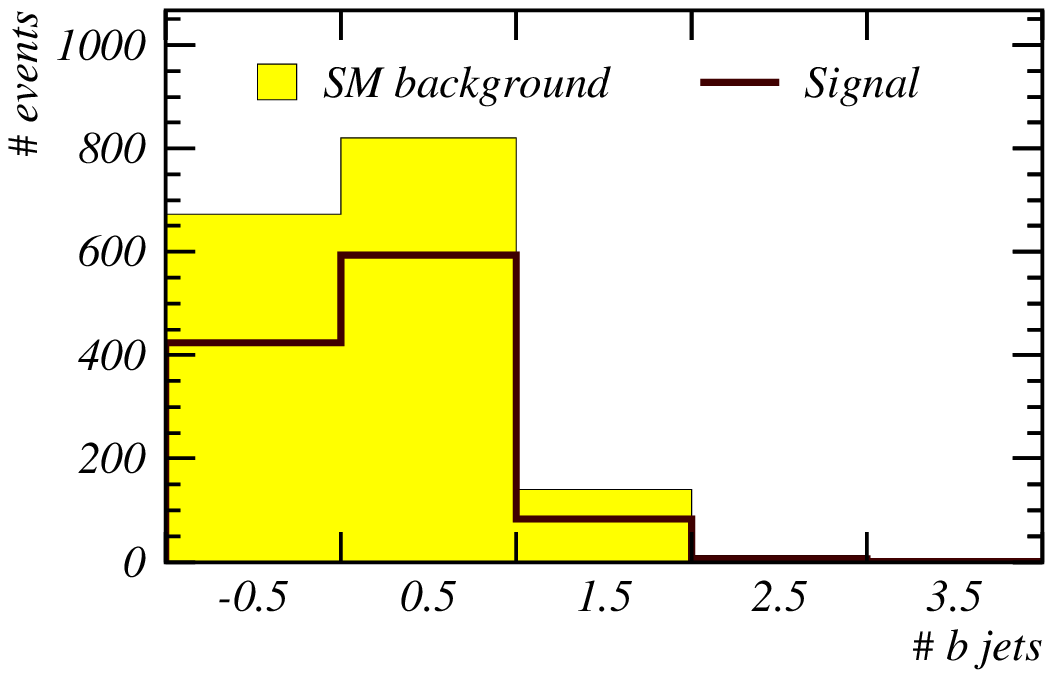}
  \vspace{-1em}

  \hspace{.24\textwidth}a)\hfill b)\hspace{.24\textwidth}\vspace{.5em}

  \includegraphics[width=.48\textwidth]{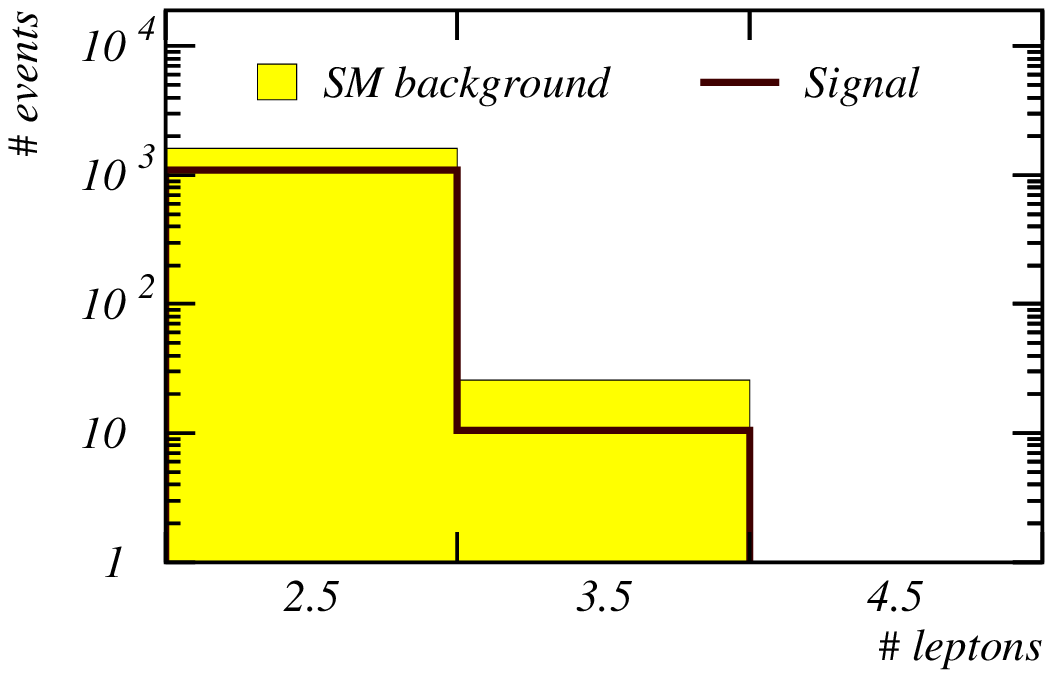}
  \hfill
  \includegraphics[width=.48\textwidth]{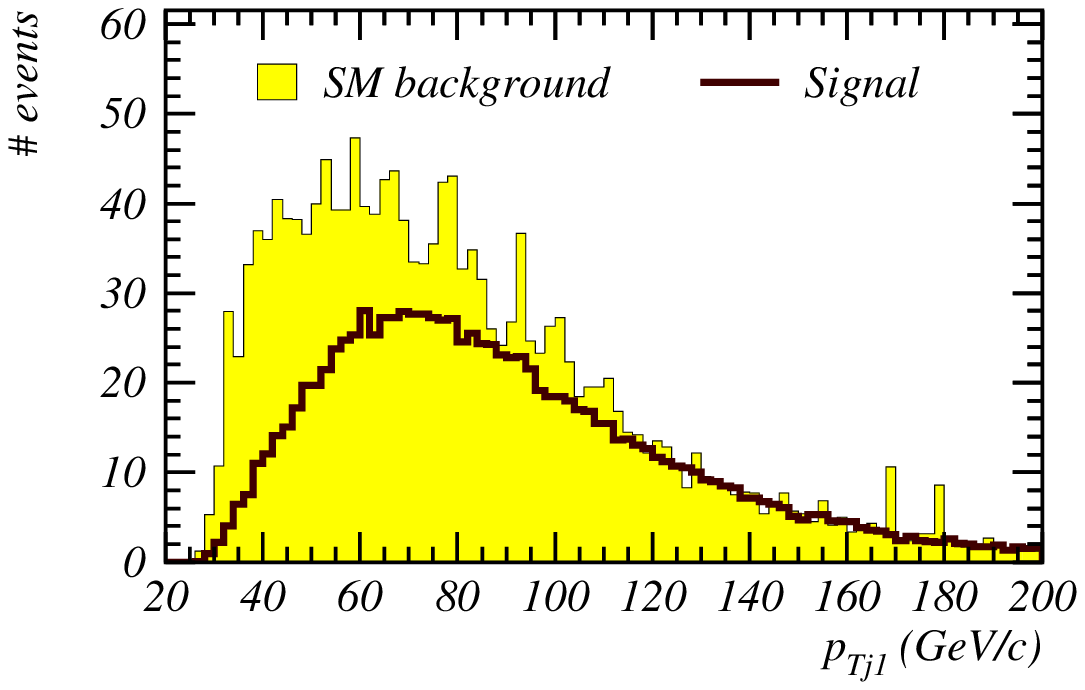}
  \vspace{-1em}

  \hspace{.24\textwidth}c)\hfill d)\hspace{.24\textwidth}\vspace{.5em}

  \includegraphics[width=.48\textwidth]{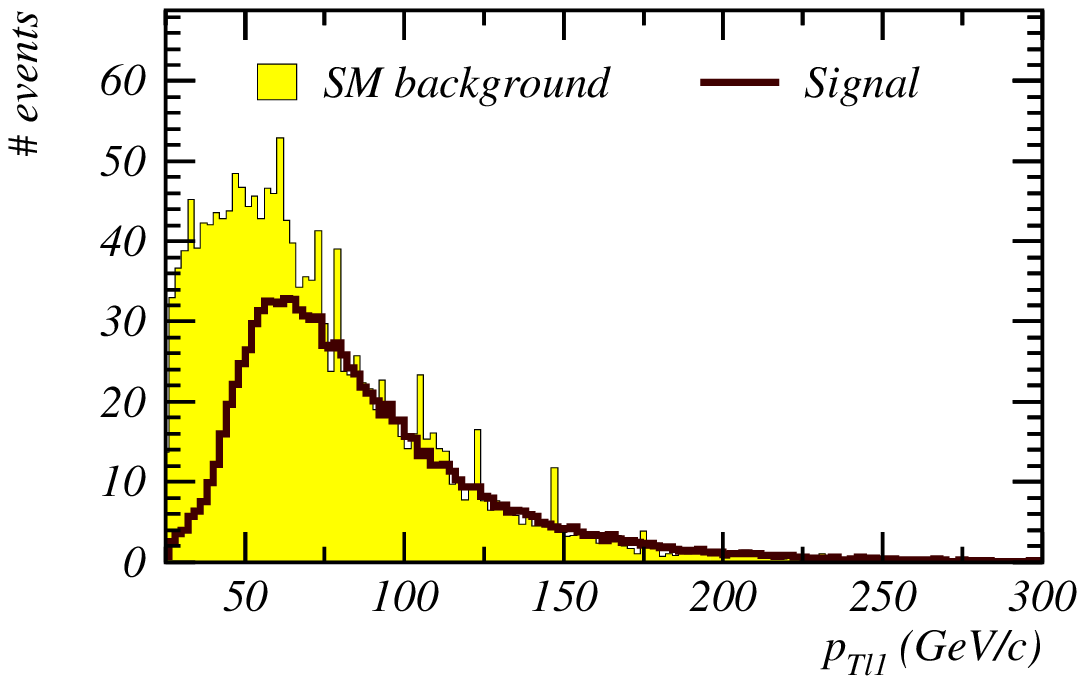}
  \hfill
  \includegraphics[width=.48\textwidth]{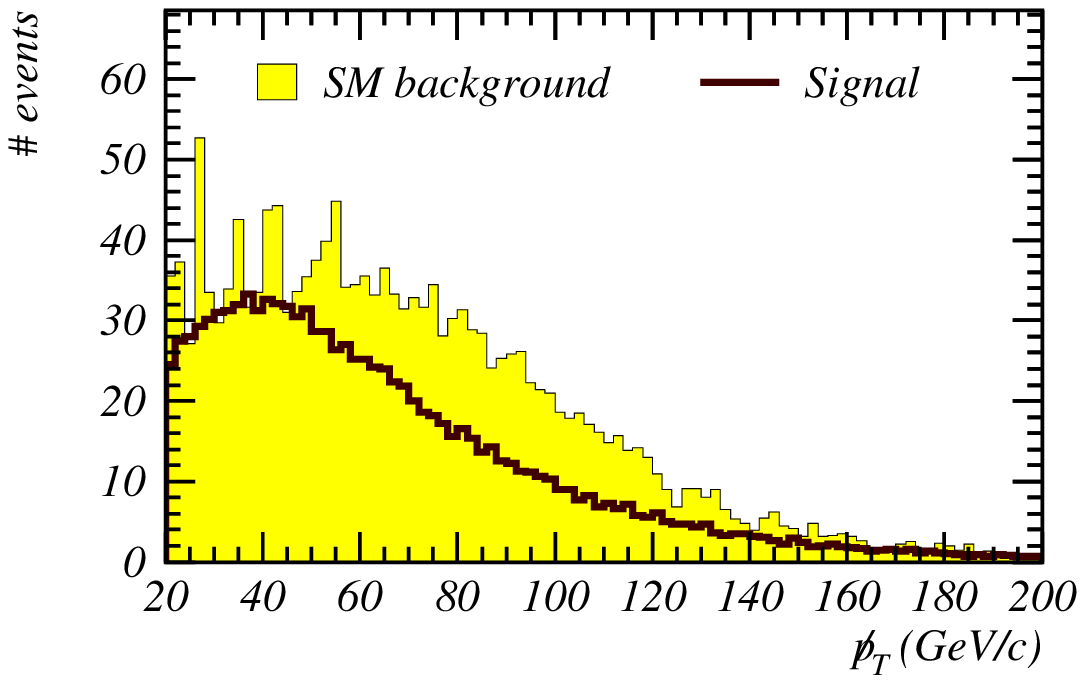}
  \vspace{-1em}

  \hspace{.24\textwidth}e)\hfill f)\hspace{.24\textwidth}

  \vspace*{-2em}
  \emph{\caption{The distributions of relevant variables for the $t\to Z
  q$ channel are shown after the preselection level:
  a)~number of jets;
  b)~number of $b$-jets;
  c)~number of leptons;
  d)~transverse momentum of the first jet;
  e)~transverse momentum of the first lepton and
  f)~missing transverse momentum.
  The SM background is normalised to $L=10$~fb$^{-1}$ and the signal 
  has an arbitrary normalization, but the same in all plots of this figure.}
  \label{fig:qz1}}
\end{figure}

At the preselection, events were accepted if they had at least two additional
isolated leptons (apart from the one already required by the general
selection criteria) with $p_T>10$~GeV$/c$ and $|\eta|<2.5$. For the 
likelihood-based analyses, all the background samples were normalised to 
$L=10$~fb$^{-1}$. The number of selected background events and the signal 
efficiency are shown in Table~\ref{tab:qz}. The distribution of relevant 
variables at this level are shown in Fig.~\ref{fig:qz1}.

\begin{table}[t]
\begin{center}
  \begin{tabular}{|l||l|l|l|l|}
    \hline
    Selection & \multirow{2}{*}{$t\rightarrow Zq$}        &  \multirow{2}{*}{$t\rightarrow\gamma q$} & \multicolumn{2}{|c|}{$t\rightarrow gq$} \\
    \cline{4-5}
    level     &                                              &                            & ``3 jets''  & ``4 jets'' \\
    \hline
    \hline
    General         & \multicolumn{4}{|c|}{1 lepton} \\
    selection       & \multicolumn{4}{|c|}{2 jets} \\
                    & \multicolumn{4}{|c|}{$\not\!p_T>20$~GeV$/c$} \\
    \hline
    Preselection    & 3 leptons                        & 1 photon                  & 3 jets                     & 4 jets                     \\
                    & 2 jets                           &                           & 1 $b$-tag                  & 1 $b$-tag                  \\
                    &                                  &                           & $E_{\mathrm{vis}}>300$~GeV & $E_{\mathrm{vis}}>300$~GeV \\
    \hline
    Final selection & ${p_T}_{j1}>30$~GeV$/c$          & ${p_T}_\gamma>75$~GeV$/c$ & ${p_T}_g>75$~GeV$/c$       & ${p_T}_g>100$~GeV$/c$      \\
                    & 1 $b$-tag                        & 1 $b$-tag                 & $m_{qg}>125$~GeV$/c$       & $m_{qg}>150$~GeV$/c$       \\
                    & 2 $\ell$ same flavour,           &                           & $m_{qg}<200$~GeV$/c$       & $m_{qg}<190$~GeV$/c$       \\
                    & \phantom{2 $\ell$} oppos. charge &                           &                            &                            \\
    \hline
  \end{tabular}
  \emph{
  \caption{Selection cuts applied to the likelihood-based analyses.}
  \label{tab:cuts}}
\end{center}
\end{table}

The final event selection was done by requiring the leading jet to have
$p_T>30$~GeV$/c$.  One of the jets had to be tagged as a $b$-jet. 
Additionally, in order to be compatible with the $Z\to \ell^+ \ell^-$ decay,
two of the three leptons present in the final state were required to have
opposite charges and the same flavour (electron or muon). The number of
selected SM events and the signal efficiency at the final selection level are
shown in Table~\ref{tab:qz}. The dominant contribution for the single top
background is the $t$-channel. No QCD ($b\bar{b}$) and $W+$jets events passed
the final selection criteria, within the generated statistics ($3.75\times
10^8$ and $3.5\times 10^7$ events, respectively).

\begin{figure}
  \includegraphics[width=.48\textwidth]{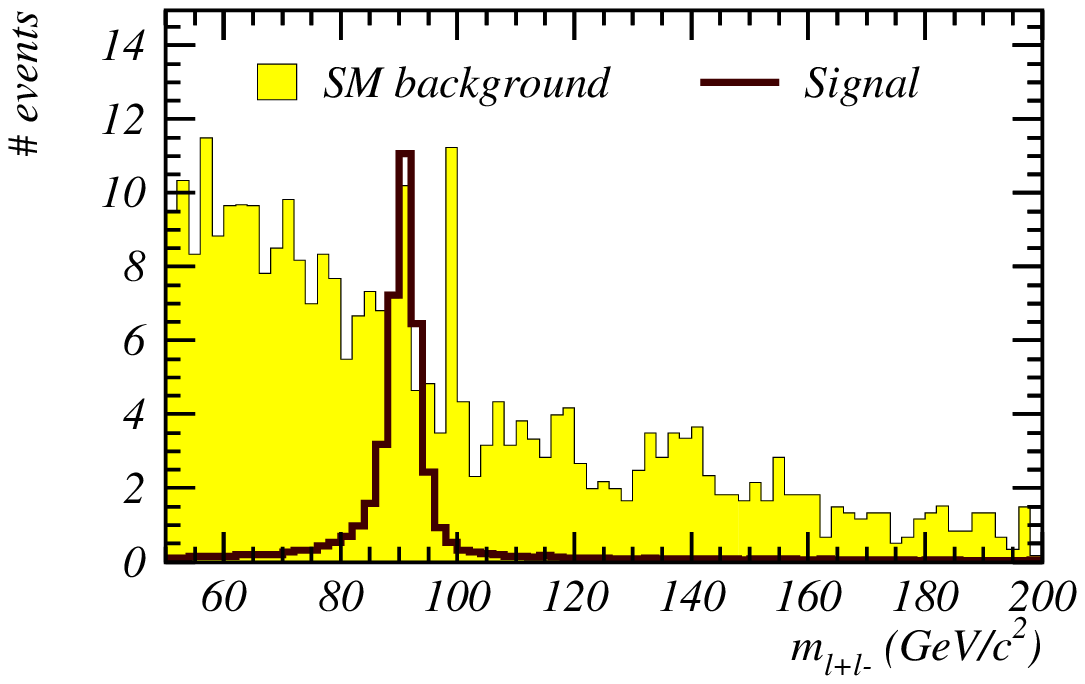}
  \hfill
  \includegraphics[width=.48\textwidth]{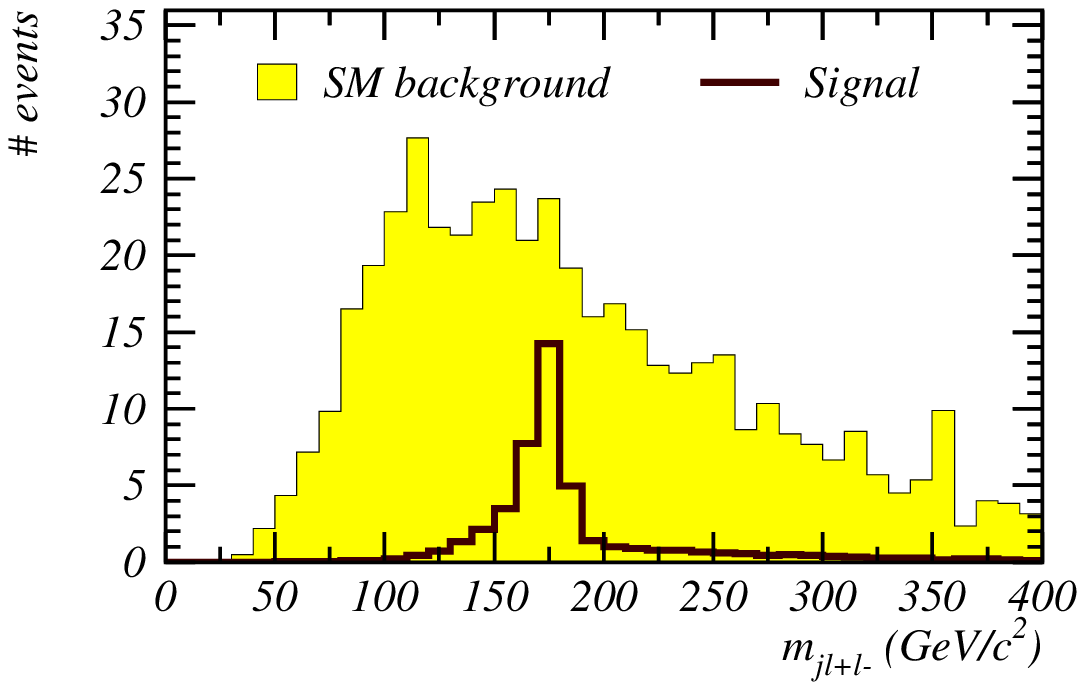}
  \vspace{-1em}

  \hspace{.24\textwidth}a)\hfill b)\hspace{.24\textwidth}\vspace{.5em}

  \includegraphics[width=.995\textwidth]{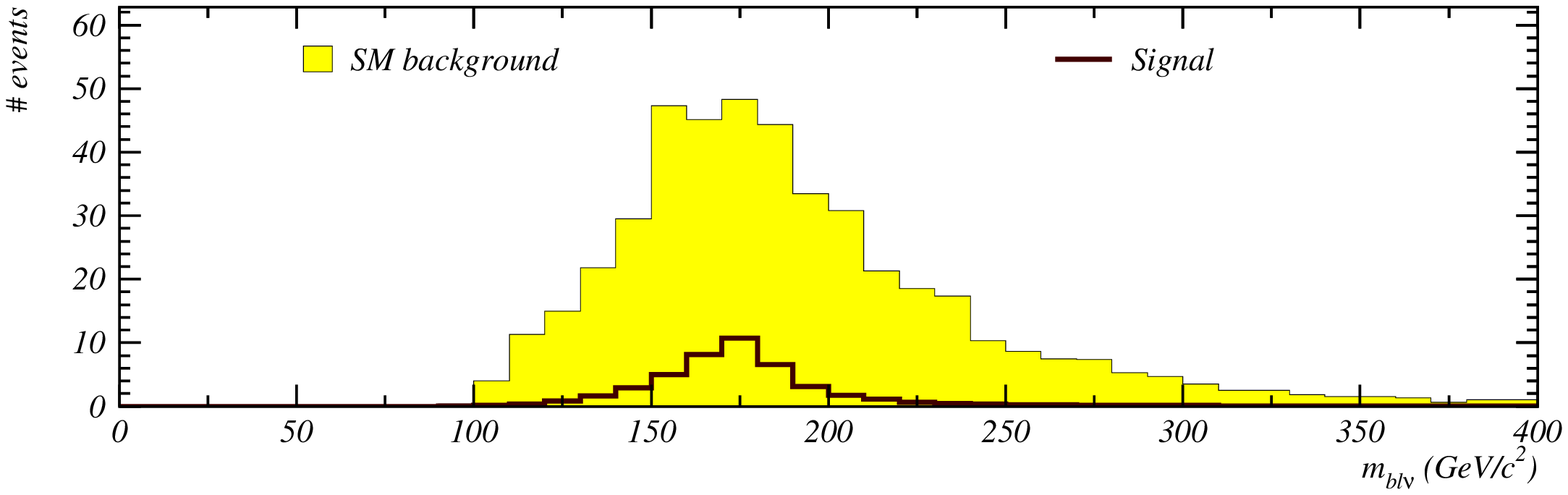}
  \vspace{-1em}

  \hspace{.48\textwidth}c)\hfill

  \vspace*{-2em}
  \emph{\caption{The reconstructed masses after the final selection level for 
  the $t\to Z q$ channel are shown:
  a)~$Z$ boson ($\ell^+ \ell^-$ invariant mass);
  b)~$t$-quark with FCNC decay ($j\ell^+ \ell^-$ invariant mass) and
  c)~$t$-quark with SM decay ($b\ell\nu$ invariant mass).
  The SM background is normalised to $L=10$~fb$^{-1}$ and the signal has
  an arbitrary normalization, but the same in all plots of this figure.}
  \label{fig:qz2}}
\end{figure}

The reconstruction of the $Z$ mass was done by calculating the $\ell^+
\ell^-$ invariant mass\footnote{In the case where more than one pair of
leptons had the same flavour and opposite charges, the pair with highest
$p_T$ was chosen.}  ($m_{\ell^+ \ell^-}$) and is shown in
Fig.~\ref{fig:qz2}a. The reconstruction of the mass of the $t$-quark which
decayed through FCNC ($m_{j\ell^+\ell^-}$) was done by associating the
non-$b$ jet with the $\ell^+ \ell^-$ pair. If more than one non-$b$ jet was
present, the one with highest $p_T$ was chosen. The signal and SM
distributions of $m_{j\ell^+ \ell^-}$ are shown in Fig.~\ref{fig:qz2}b. The
decay of the other $t$-quark ($t\to b\ell\nu$) cannot be directly
reconstructed due to the presence of an undetected neutrino in the final
state. Nevertheless, the neutrino four-momentum can be estimated by assuming
the transverse missing energy to be the transverse neutrino momentum. The
longitudinal component can be determined, with a quadratic ambiguity, by
constraining the $W$ mass (calculated as the invariant mass of the neutrino
and the most energetic remaining lepton) to its known central value
($m_W=80.4$~GeV$/c^2$~\cite{pdg}). The mass of $t$-quark with a SM decay,
shown in Fig.~\ref{fig:qz2}c, was reconstructed by associating the $b$-jet
with the reconstructed $W$. The quadratic ambiguity was removed by choosing
the solution closer to $m_t=175$~GeV$/c^2$.

Following the final selection, a likelihood-based type of analysis was
applied. Signal ($\mathcal P^{signal}_i$) and background-like ($\mathcal
P^{back.}_i$) probabilities were computed using Probability Density
Functions (p.d.f.), constructed from relevant physical variables. The
signal $\mathcal L_S=\Pi^{n}_{i=1}\mathcal P^{signal}_i$ and background
$\mathcal L_B=\Pi^{n}_{i=1}\mathcal P^{back.}_i$ likelihoods ($n$ is the
number of p.d.f.) were used to built the discriminant variable, defined
as $L_R=\ln\left({\mathcal L}_S/{\mathcal L}_B\right)$, assuming uncorrelated 
variables.

\begin{figure}
  \includegraphics[width=.48\textwidth]{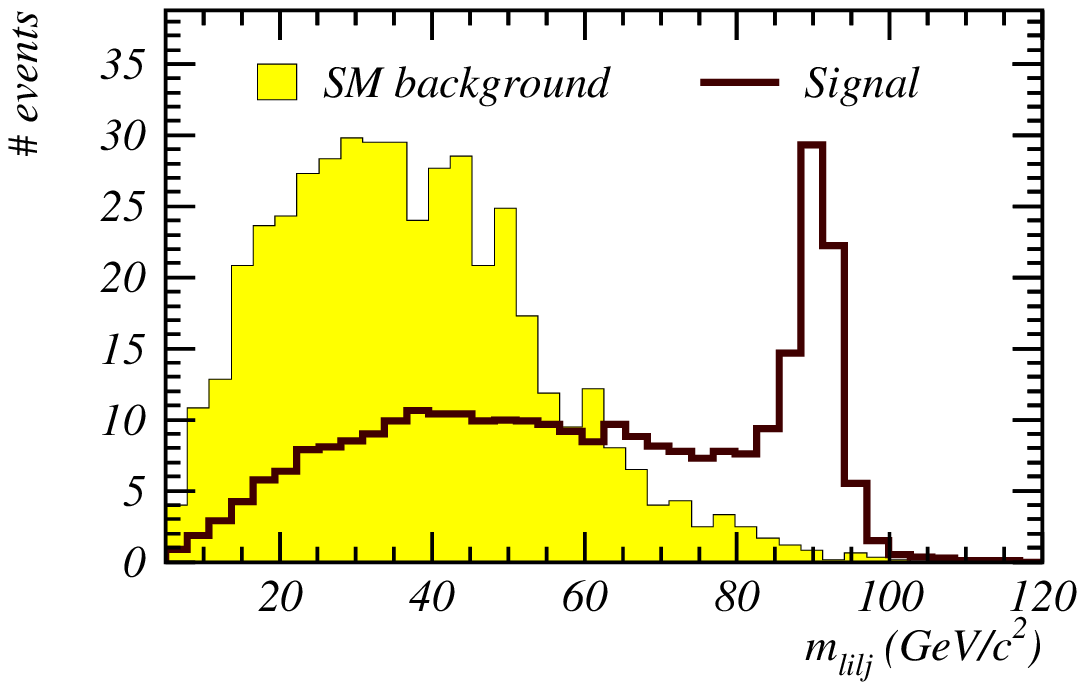}
  \hfill
  \includegraphics[width=.48\textwidth]{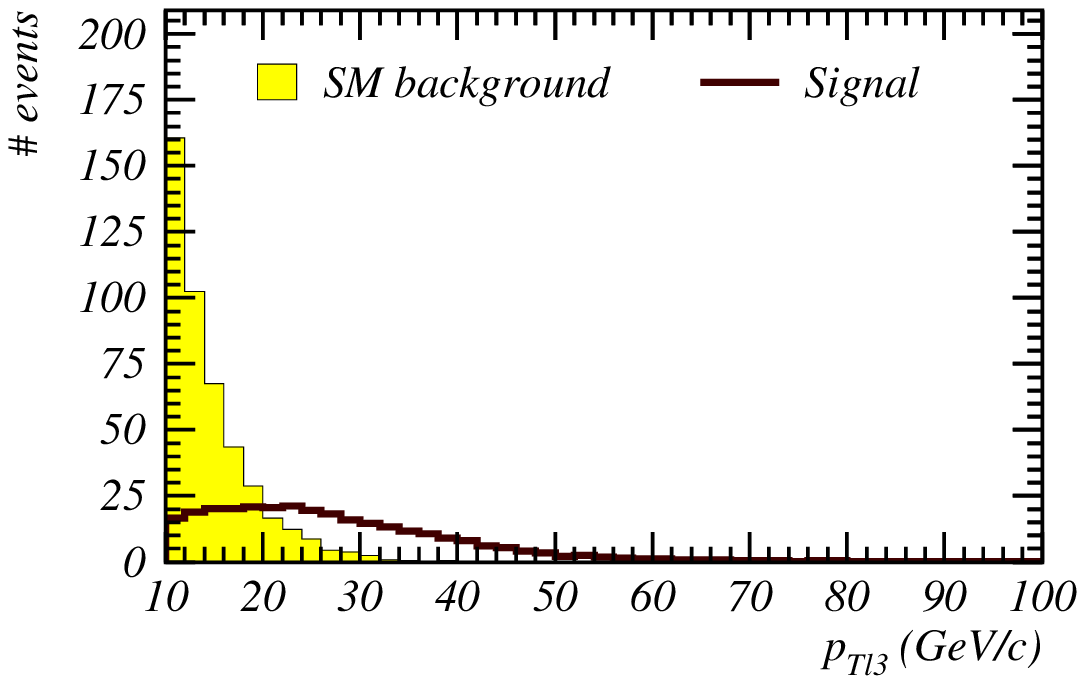}
  \vspace{-1em}

  \hspace{.24\textwidth}a)\hfill b)\hspace{.24\textwidth}\vspace{.5em}  

  \includegraphics[width=.995\textwidth]{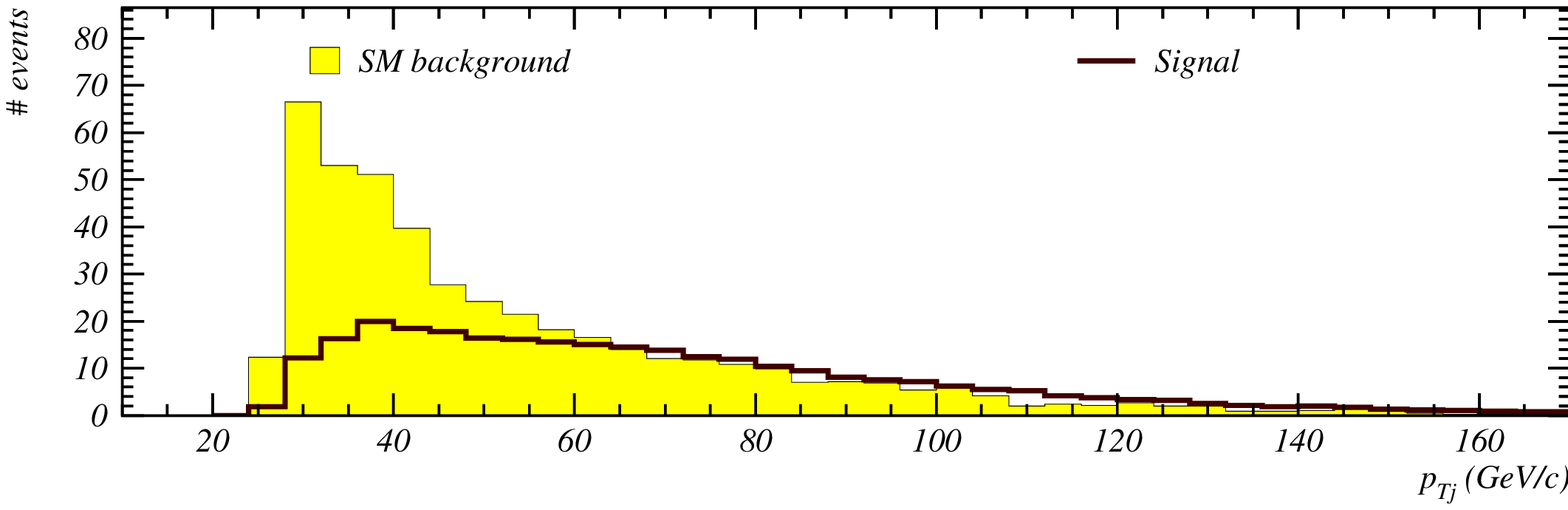}
  \vspace{-1em}

  \hspace{.48\textwidth}c)\hfill

  \vspace*{-1.5em}
  \emph{\caption{The distribution of the variables based on which the 
  p.d.f. were built are shown ($t\to Z q$ channel):
  a)~two leptons minimum mass (only the first three leptons were 
  considered);
  b)~transverse momentum of the third lepton and
  c)~transverse momentum of the most energetic non $b$-jet.
  The $j\ell^+\ell^-$ invariant mass was also used as p.d.f. and is shown in 
  Fig.~\ref{fig:qz2}b.
  The SM background is normalised to $L=10$~fb$^{-1}$ and the signal has
  an arbitrary normalization, but the same in all plots of this figure.}
  \label{fig:qz3}}
\end{figure}

For the $t\to Z q$ channel the p.d.f. were based on the following physical 
distributions (\emph{c.f.} Fig.~\ref{fig:qz3}):  
\begin{itemize} 
\item minimum invariant mass ($m_{\ell_i\ell_j}$) of the three possible 
  combinations of two leptons (only the three leading leptons were 
  considered);
\item transverse momentum of the third lepton (${p_T}_{l3}$);  
\item the $j\ell^+ \ell^-$ invariant mass and 
\item the transverse momentum of the most
  energetic non-$b$ jet (${p_T}_j$).  
\end{itemize}

\begin{figure}
  \begin{center}
    \includegraphics[width=.995\textwidth]{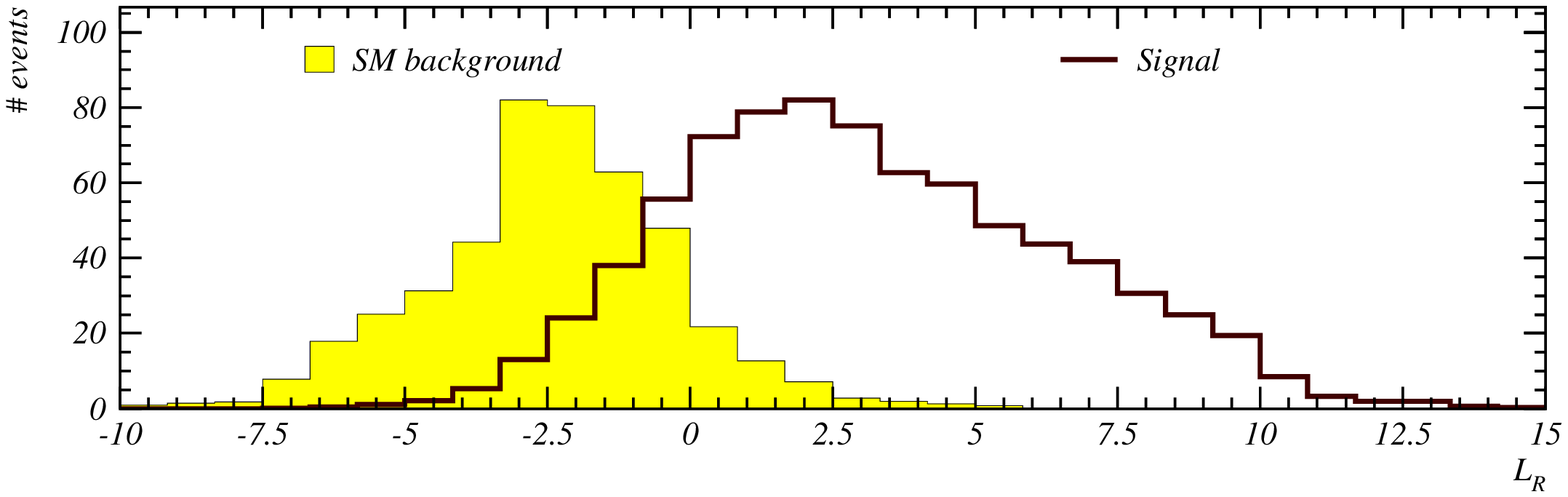}
  \end{center}

  \vspace*{-3em}
  \emph{\caption{SM background and signal discriminant variable distributions for 
  the $t\to Z q$ channel are shown. The SM background is normalised to 
  $L=10$~fb$^{-1}$ and the signal has
  an arbitrary normalization.}
  \label{fig:qz4}}
\end{figure}

\begin{figure}
  \begin{center}
    \includegraphics[width=.995\textwidth]{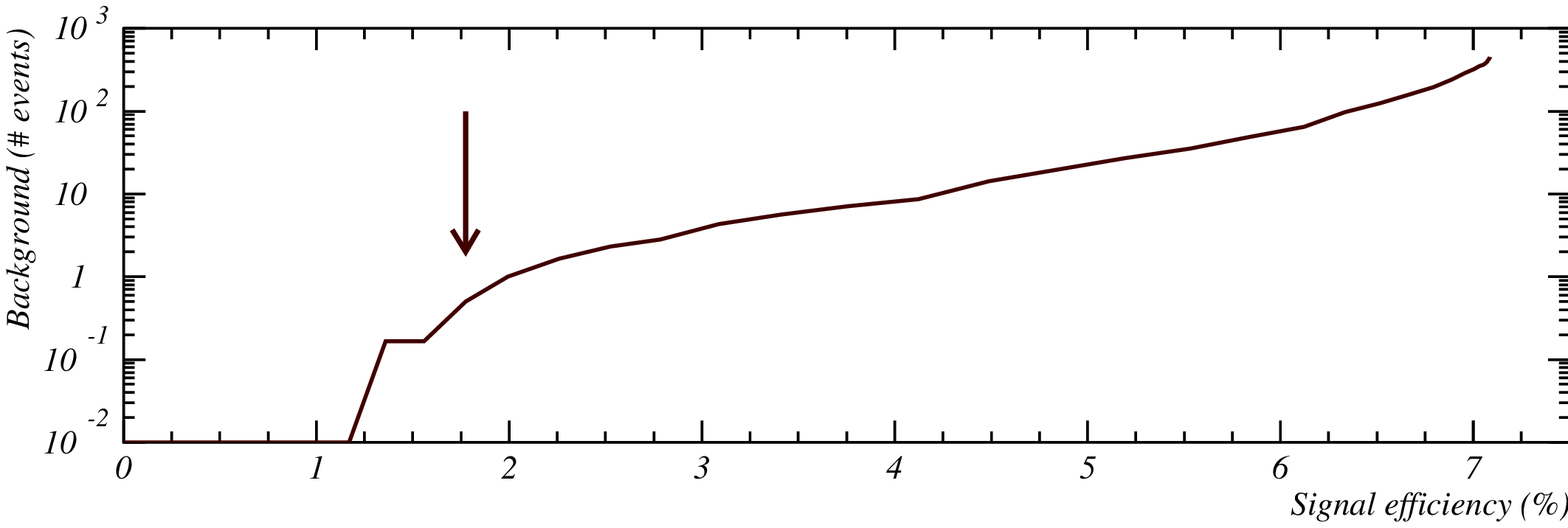}
  \end{center}

  \vspace*{-3em}
  \emph{\caption{The number of expected SM background as a function of 
  the signal efficiency for the $t\to Z q$ channel is shown.
  The SM background is normalised to $L=10$~fb$^{-1}$.
  The arrow shows the point with best $S/\sqrt{B}$.}
  \label{fig:qz:effvsback}}
\end{figure}

The $L_R$ distributions for SM background and signal are shown, after the
final selection, in Fig.~\ref{fig:qz4}. The number of expected SM background
as a function of the signal efficiency obtained by cutting the discriminant
variable is shown in Fig.~\ref{fig:qz:effvsback}.

\begin{table}[t]
\begin{center}
  \begin{tabular}{|l||r||r|r|r|r|}
    \hline
    \multicolumn{1}{|c||}{Description} & \multicolumn{1}{|c||}{Signal}             & \multicolumn{4}{|c|}{Background Processes}\\
    \cline{3-6}
    \multicolumn{1}{|c||}{of}          & \multicolumn{1}{|c||}{$t\rightarrow Zq$}  & \multicolumn{1}{|c|}{$Z+$jets} & \multicolumn{1}{|c|}{$Z+W$} & \multicolumn{1}{|c|}{$t\bar t$}  & \multicolumn{1}{|c|}{single $t$}\\
    \cline{2-6}
    \multicolumn{1}{|c||}{Cuts}        & \multicolumn{1}{|c||}{$\varepsilon$ (\%)} & \multicolumn{1}{|c|}{Nevt} & \multicolumn{1}{|c|}{Nevt} & \multicolumn{1}{|c|}{Nevt} & \multicolumn{1}{|c|}{Nevt} \\
    \hline
    \hline
    Preselection    & 17.0 & 78.7 & 29.8 &  1514.2 & 24.3 \\
    \hline
    Final Selection &  7.1 & 13.1 &  1.7 &   434.2 &  4.8 \\
    \hline
  \end{tabular}
  \emph{
  \caption{The number of selected background events, normalised to 
  $L=10$~fb$^{-1}$, and signal efficiencies in the $t\to Zq$
  channel for the preselection and final selection levels, obtained 
  with a likelihood-based analysis, are shown. }
  \label{tab:qz}}
\end{center}
\end{table}

\subsection{$\boldsymbol{t\to \gamma q}$ channel}

The $t\bar t$ final states corresponding to the FCNC decay $t\to \gamma
q$ are characterised by the presence of a high $p_T$ photon and a
light jet from the top quark decay. Since the existence of the photon is
not sufficient to reduce the QCD background, only the leptonic decays of
the $W$ (originated from the SM decay of the other top quark) were
considered. The final states corresponding to these signal events are
characterised by a topology with two jets (one $b$-jet from the SM top
decay), one high $p_T$ photon, one lepton and missing transverse
momentum from the undetected neutrino. Following a previous
analysis~\cite{dodd}, a new one was developed~\cite{carvalho} and is
described below.

\begin{figure}
  \psfrag{0.5}{\small 1}
  \psfrag{-0.5}{\small 0}
  \psfrag{1.5}{\small 2}
  \psfrag{2.5}{\small 3}
  \psfrag{3.5}{\small 4}
  \psfrag{4.5}{\small 5}
  \psfrag{5.5}{\small 6}
  \psfrag{6.5}{\small 7}
  \psfrag{7.5}{\small 8}
  \psfrag{8.5}{\small 9}

  \includegraphics[width=.48\textwidth]{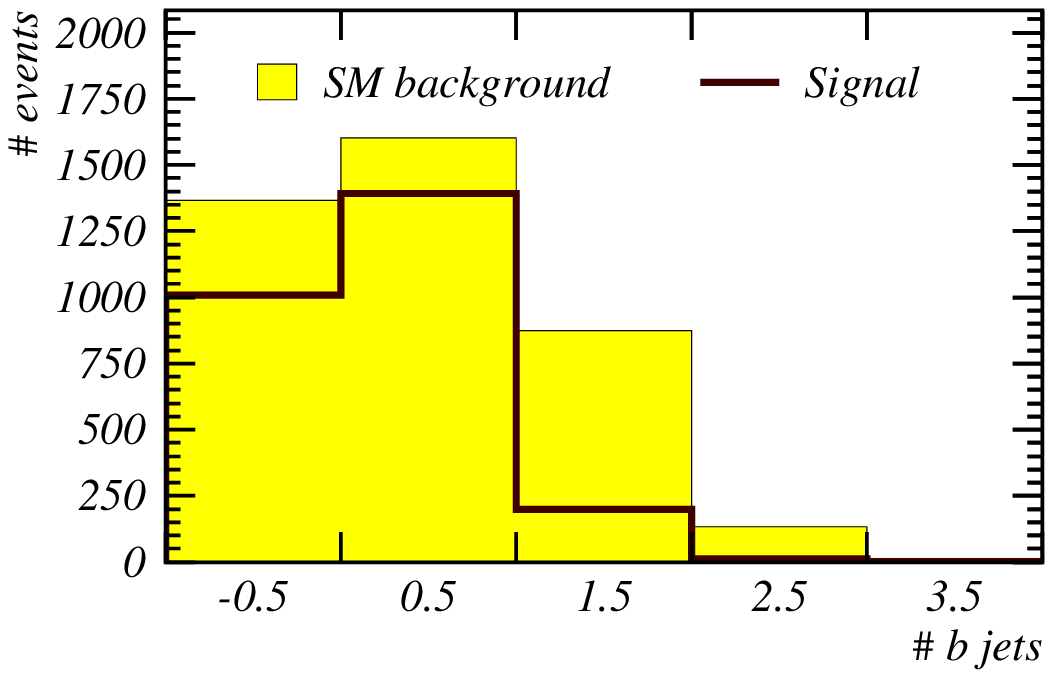}
  \hfill
  \includegraphics[width=.48\textwidth]{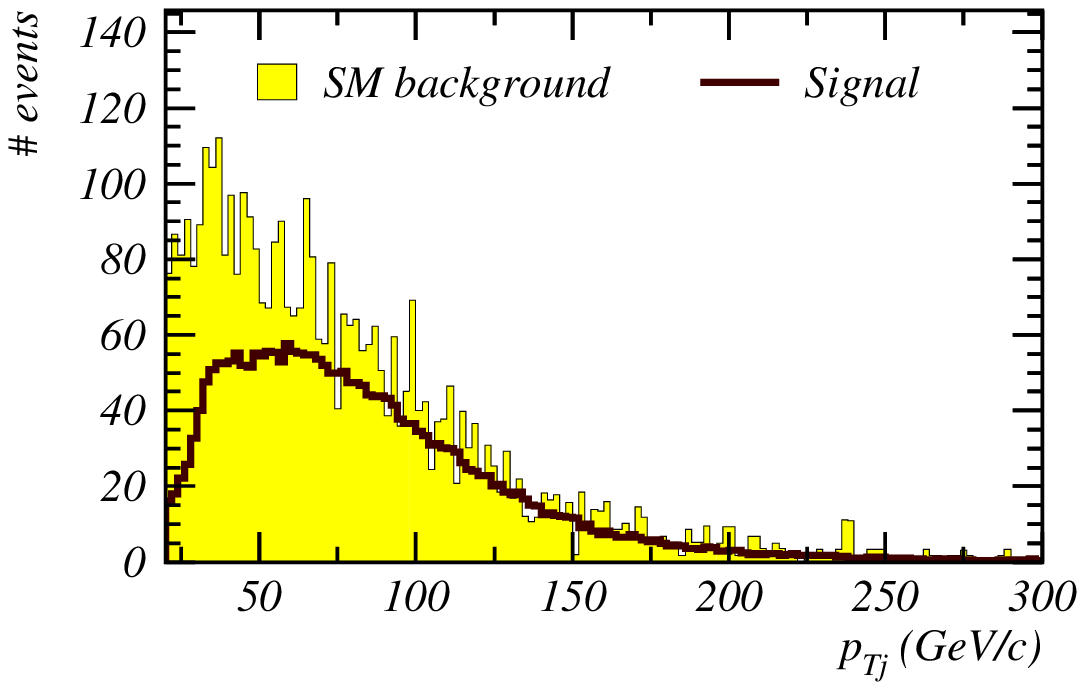}
  \vspace{-1em}

  \hspace{.24\textwidth}a)\hfill b)\hspace{.24\textwidth}

%

  \vspace*{-2em}
  \emph{\caption{The distributions of relevant variables for the $t\to
  \gamma q$ channel are shown after the preselection level:
  a)~number of $b$-jets and
  b)~transverse momentum of the $c(u)$-jet.
  The SM background is normalised to $L=10$~fb$^{-1}$ and the signal has
  an arbitrary normalization, but the same in all plots of this figure.}
  \label{fig:qgamma1}}
\end{figure}

After applying the general selection criteria described in
section~\ref{sec:qz}, a preselection was defined by requiring the events
to have at least one photon with $p_T>50$~GeV$/c$ and $|\eta|<2.5$.
Additionally, in order to prevent events to be simultaneously assigned
to the $t\to Z q$ and $t\to \gamma q$ channels, a maximum of two
leptons in the final state were allowed. The number of selected SM
background events and the signal efficiency at the preselection level
are shown in Table~\ref{tab:qgamma}. The distributions of relevant
variables for SM background and signal are shown in
Fig.~\ref{fig:qgamma1}.

\begin{table}[b
]
\begin{center}
  \begin{tabular}{|l||r||r|r|r|r|}
    \hline
    \multicolumn{1}{|c||}{Description} & \multicolumn{1}{|c||}{Signal}             & \multicolumn{4}{|c|}{Background Processes}\\
    \cline{3-6}
    \multicolumn{1}{|c||}{of}          & \multicolumn{1}{|c||}{$t\rightarrow\gamma q$}  & \multicolumn{1}{|c|}{$Z+$jets} & \multicolumn{1}{|c|}{$Z+W$} & \multicolumn{1}{|c|}{$t\bar t$}  & \multicolumn{1}{|c|}{single $t$}\\
    \cline{2-6}
    \multicolumn{1}{|c||}{Cuts}        & \multicolumn{1}{|c||}{$\varepsilon$ (\%)} & \multicolumn{1}{|c|}{Nevt} & \multicolumn{1}{|c|}{Nevt} & \multicolumn{1}{|c|}{Nevt} & \multicolumn{1}{|c|}{Nevt} \\
    \hline
    \hline
    Preselection    & 23.3 & 584.2 & 325.7 & 2832.4 & 206.2 \\
    \hline
    Final Selection &  6.9 &  15.2 &   7.7 &  271.6 &  23.0 \\
    \hline
  \end{tabular}
  \emph{
  \caption{The number of selected background events, normalised to 
  $L=10$~fb$^{-1}$, and signal efficiencies
  in the $t\to\gamma q$ channel
  for the preselection and final selection levels, obtained with a 
  likelihood-based analysis, are shown. }
  \label{tab:qgamma}}
\end{center}
\end{table}

\begin{figure}
  \psfrag{0.5}{\small 1}
  \psfrag{-0.5}{\small 0}
  \psfrag{1.5}{\small 2}
  \psfrag{2.5}{\small 3}
  \psfrag{3.5}{\small 4}
  \psfrag{4.5}{\small 5}
  \psfrag{5.5}{\small 6}
  \psfrag{6.5}{\small 7}
  \psfrag{7.5}{\small 8}
  \psfrag{8.5}{\small 9}

  \includegraphics[width=.48\textwidth]{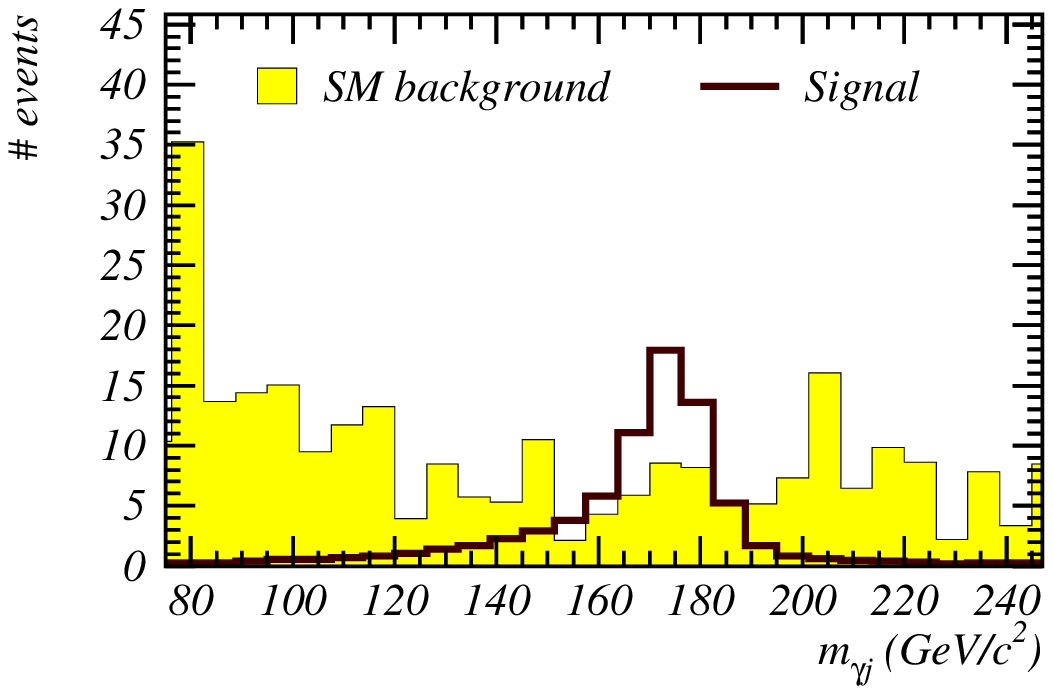}
  \hfill
  \includegraphics[width=.48\textwidth]{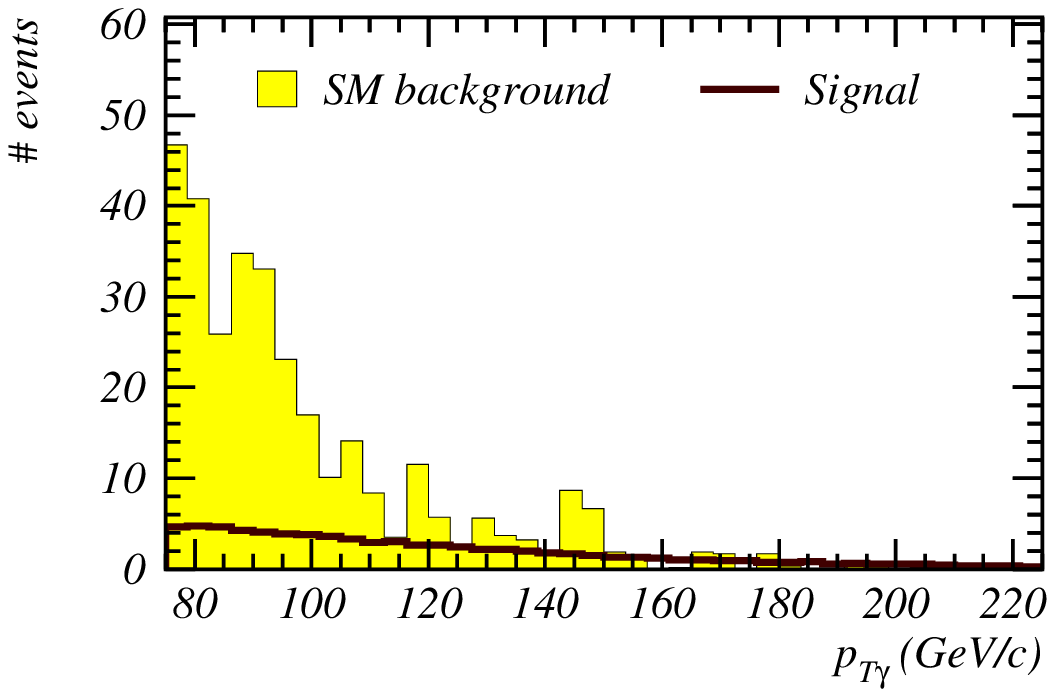}
  \vspace{-1em}

  \hspace{.24\textwidth}a)\hfill b)\hspace{.24\textwidth}\vspace{.5em}  

  \includegraphics[width=.995\textwidth]{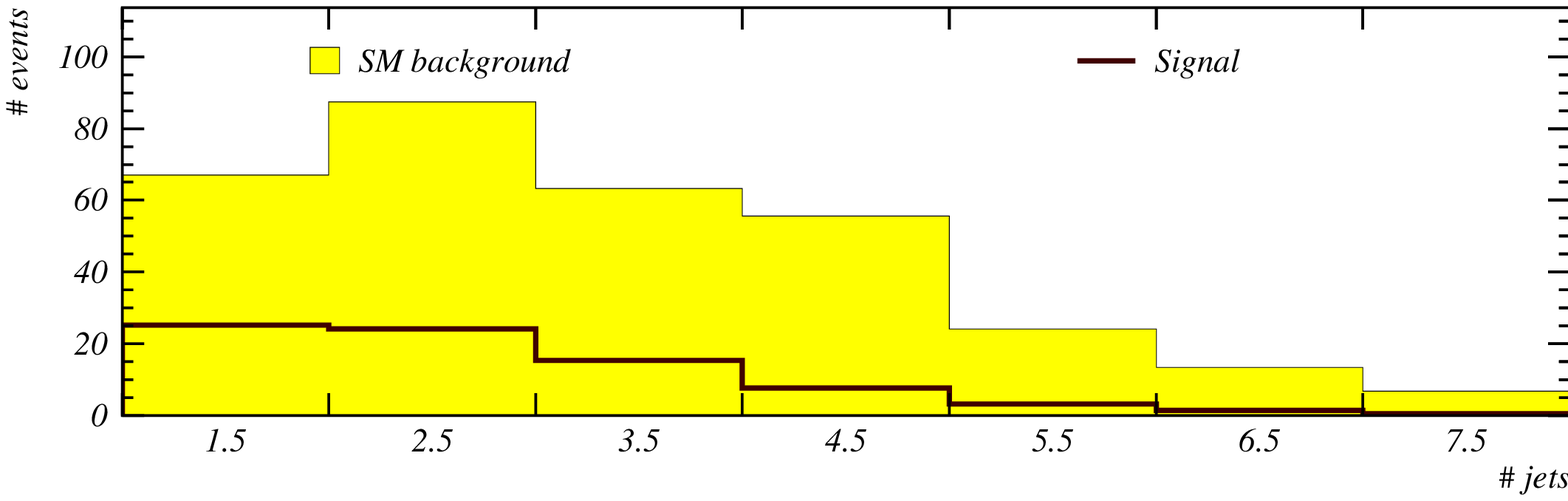}
  \vspace{-1em}

  \hspace{.48\textwidth}c)\hfill

  \vspace*{-1.5em}
  \emph{\caption{The distribution of the variables based on which the
  p.d.f. were built are shown ($t\to \gamma q$ channel):
  a)~reconstructed mass of the $t$-quark with FCNC decay ($j\gamma$ invariant 
  mass);
  b)~transverse momentum of the photon and
  c)~number of jets.
  The SM background is normalised to $L=10$~fb$^{-1}$ and the signal has
  an arbitrary normalization, but the same in all plots of this figure.}
  \label{fig:qgamma2}}
\end{figure}

The final event selection was done by requiring the leading photon to have
$p_T>75$~GeV$/c$ and one of the two jets with highest $p_T$ to be tagged as a
$b$-jet (only one $b$-jet was allowed). This selection largely reduces the
background, being $t\bar t_{SM}$ the dominant contribution, as can be seen in
Table~\ref{tab:qgamma}. The FCNC top decay was reconstructed using the
non-$b$ jet and the photon (in the cases were more than one photon or non-$b$
quark were available, the one with higher $p_T$ was chosen). No QCD
($b\bar{b}$) and $W+$jets events, within the generated statistics, passed the
final selection criteria. Just like for the $t\to Z q$ channel, a
likelihood-based type of analysis was used. The p.d.f. were built based on
the following variables (\emph{c.f.}  Fig.~\ref{fig:qgamma2}):
\begin{itemize}
  \item the mass of the $t$-quark with FCNC decay, reconstructed from the 
  photon and the non-$b$ jet ($m_{j\gamma}$);
  \item the transverse momentum of the leading photon (${p_T}_\gamma$) and
  \item the number of jets.
\end{itemize}

\begin{figure}
  \begin{center}
    \includegraphics[width=.995\textwidth]{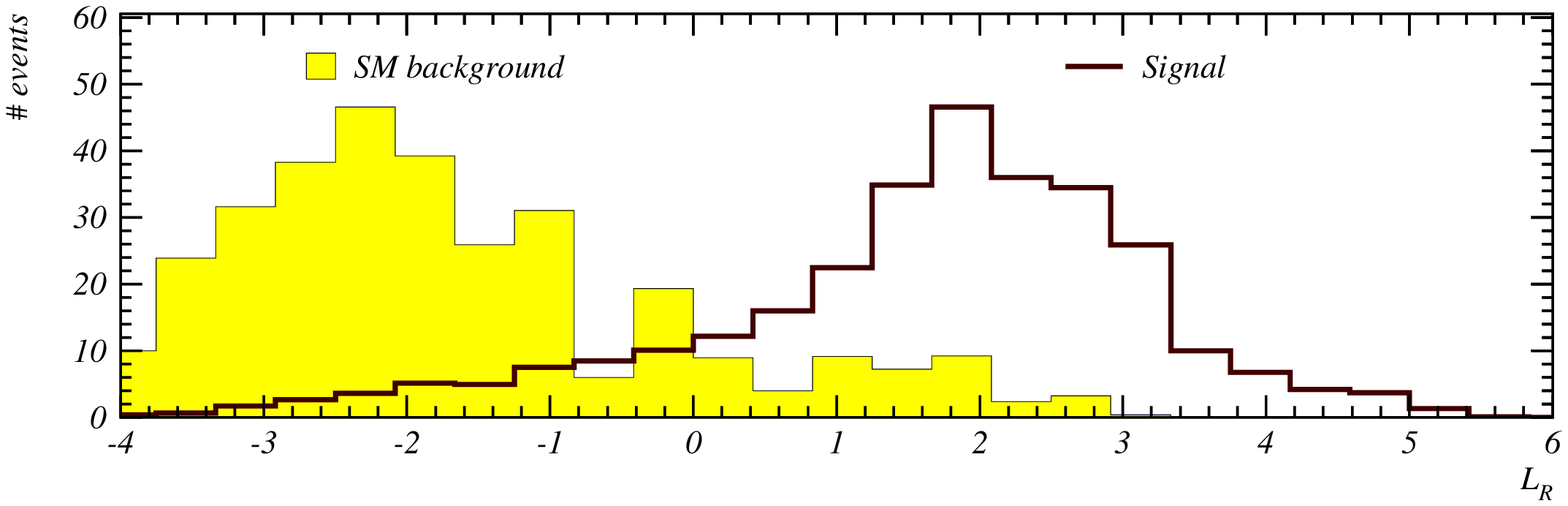}
  \end{center}

  \vspace*{-3em}
  \emph{\caption{SM background and signal discriminant variable distributions for
  the $t\to \gamma q$ channel are shown. The SM background is normalised to
  $L=10$~fb$^{-1}$ and the signal has
  an arbitrary normalization.}
  \label{fig:qgamma3}}
\end{figure}

\begin{figure}
  \begin{center}
    \includegraphics[width=.995\textwidth]{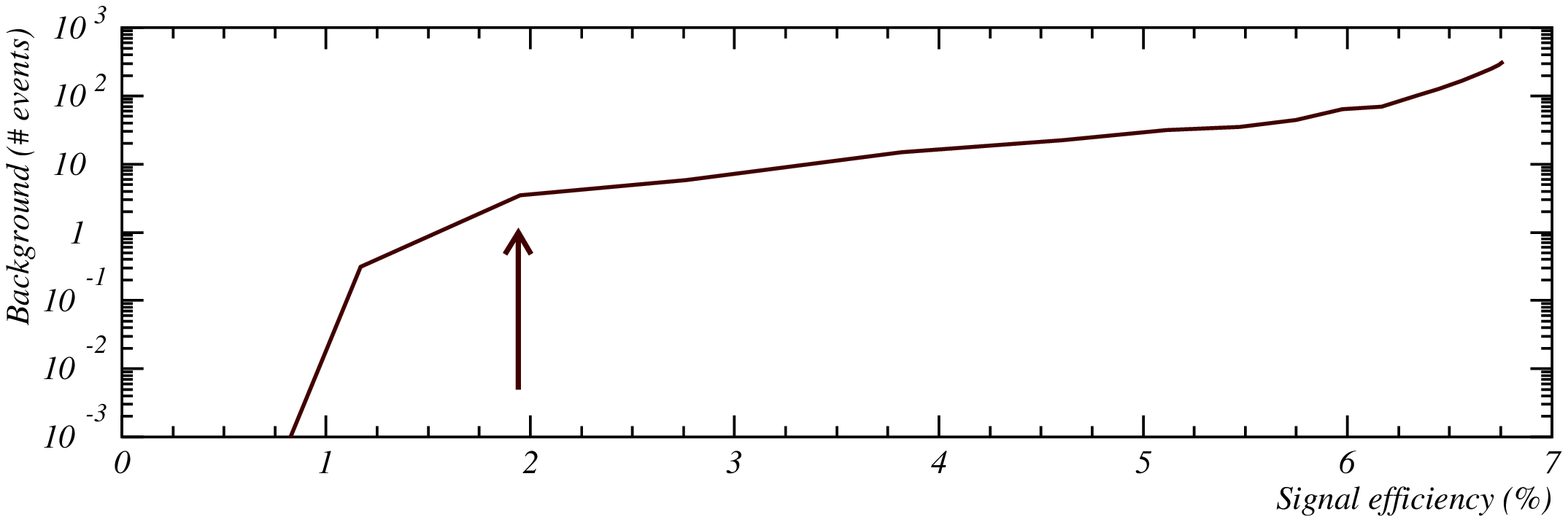}
  \end{center}

  \vspace*{-3em}
  \emph{\caption{The number of expected SM background as a function of 
  the signal efficiency for the $t\to\gamma q$ channel is shown.
  The SM background is normalised to $L=10$~fb$^{-1}$.
  The arrow shows the point with best $S/\sqrt{B}$.}
  \label{fig:qgamma:effvsback}}
\end{figure}

The discriminant variables distributions for signal and SM expectation are
shown in Fig.~\ref{fig:qgamma3} and the number of expected SM background as a
function of the signal efficiency obtained by cutting the discriminant
variable is shown in Fig.~\ref{fig:qgamma:effvsback}

\subsection{$\boldsymbol{t\to g q}$ channel}

The final states of $t\bar t$ events with one of the top quarks decaying
into a gluon, $t\to gq$, are characterised by the presence of a high
$p_T$ gluon and a light jet from the top quark decay. Only the leptonic
decays of the $W$ (originated from the SM decay of the other top quark)
were taken into account, otherwise the final state would be fully
hadronic and the signal would be overwhelmed by the QCD background. The
final states are thus characterised by the existence of at least three
jets (one $b$-jet from the SM top decay), one lepton and missing
transverse momentum from the undetected neutrino.

Although no previous analyses have been performed for the $t\to g q$
decay, the anomalous coupling $tgq$ in top production was studied
in the past~\cite{ATLAS,cakir}. A new analysis dedicated to 
the $t\to g q$ decay was developed~\cite{carvalho} and is described 
here.

As in this topology the FCNC top decay corresponds to a fully hadronic
final state, a more restrictive event selection was necessary. The
general selection criteria of section~\ref{sec:qz} was applied to the
events. At the preselection, events were required to have only one
lepton and no photons with transverse momentum above $p_T>5$~GeV$/c$, to
reject events assigned to the other FCNC channels. The total visible
energy ($E_{\mathrm{vis}}$) had to be greater than 300~GeV. At least three jets with
$|\eta|<2.5$ and $p_T>20$~GeV$/c$ were required. For the leading jet the
cut was increased to $40$~GeV$/c$. The events were then classified as
``3 jets'' or ``4 jets'' if they had exactly three jets or at least 4
jets, respectively. 

\subsubsection{The ``3 jets'' sample}

\begin{figure}
  \psfrag{mat}[rt][rt]{\small $m_{\bar t}$ (GeV/$c^2$)}
  \psfrag{ptat}[rt][rt]{\small ${p_T}_{\bar t}$ (GeV/$c^2$)}

  \includegraphics[width=.48\textwidth]{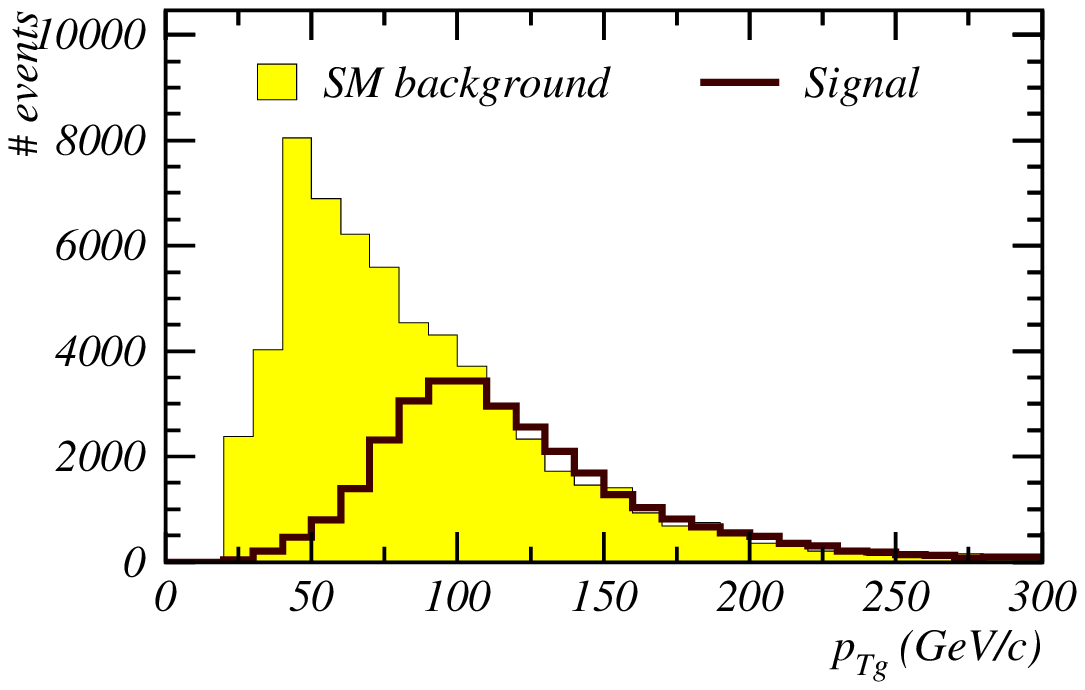}
  \hfill
  \includegraphics[width=.48\textwidth]{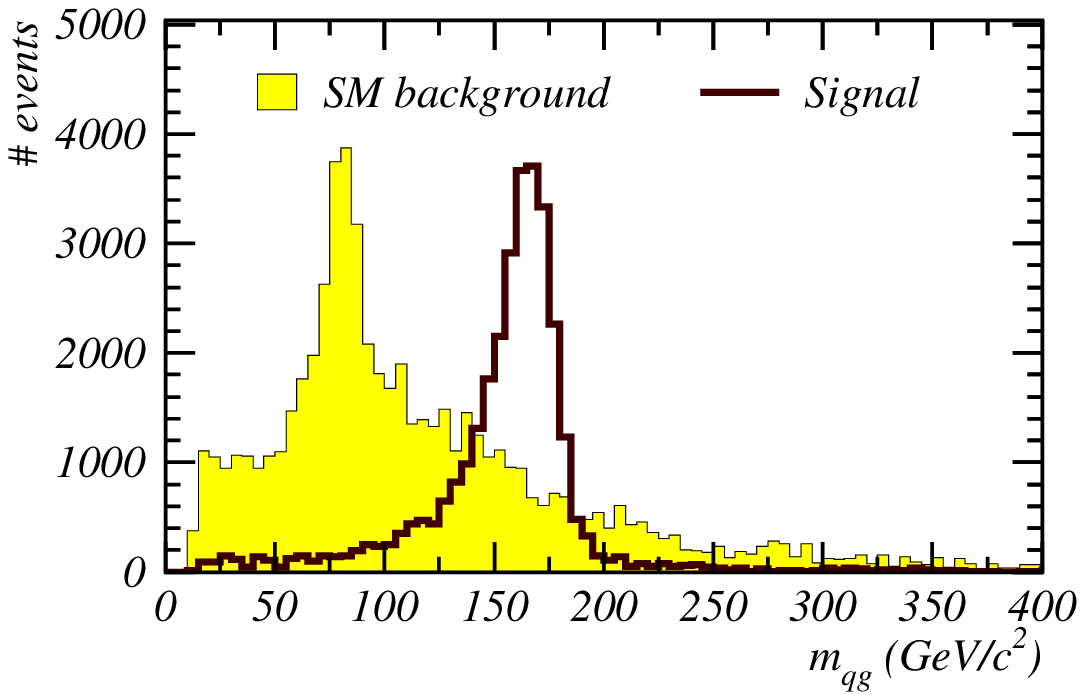}
  \vspace{-1em}

  \hspace{.24\textwidth}a)\hfill b)\hspace{.24\textwidth}\vspace{.5em}

  \includegraphics[width=.995\textwidth]{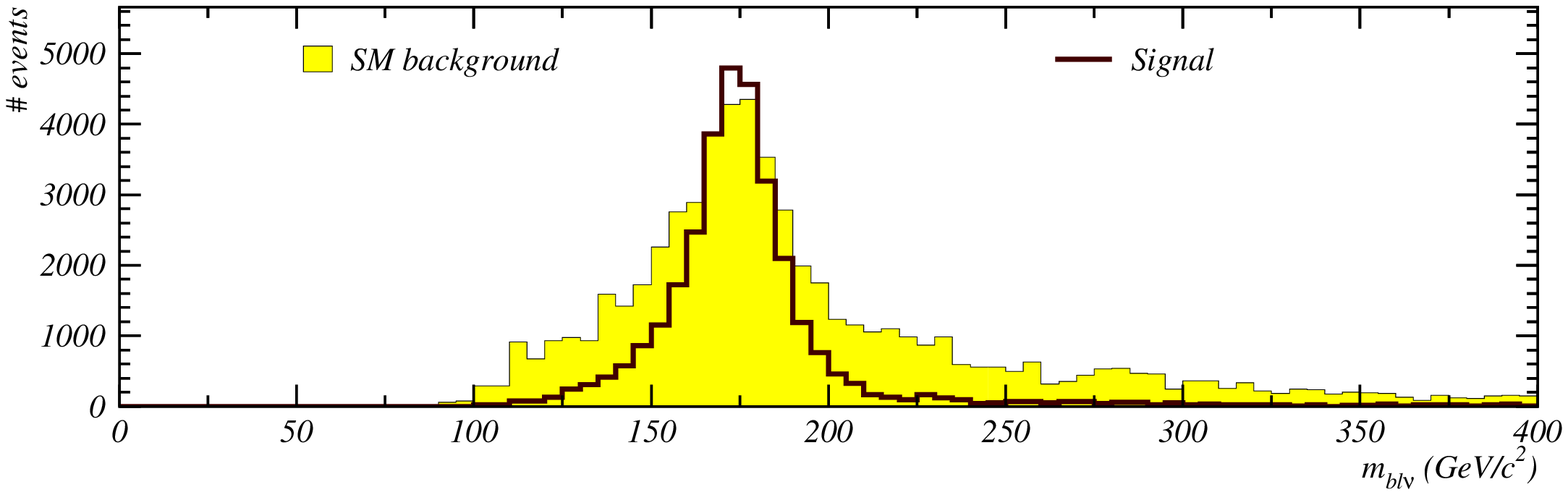}
  \vspace{-1em}

  \hspace{.48\textwidth}c)\hfill

  \vspace*{-2em}
  \emph{\caption{The distributions of relevant variables for the $t\to g
  q$ (``3 jets'') channel are shown after the preselection level:
  a)~transverse momentum of the gluon;
  b)~the $qg$ invariant mass and
  c)~the $b\ell\nu$ invariant mass.
  The SM background is normalised to $L=10$~fb$^{-1}$ and the signal has 
  an arbitrary normalization, but the same in all plots of this figure.}
  \label{fig:qg3j1}}
\end{figure}

The preselection was completed by requiring only one $b$-tagged jet in the
event. The gluon jet was assumed to be the non-$b$ jet with the highest
transverse momentum. This distribution is shown in Fig.~\ref{fig:qg3j1},
together with the mass of the $t$-quark with FCNC decay ($m_{qg}$),
reconstructed from the non-$b$ jets. The mass of the $t$-quark with SM decay
($m_{b\ell\nu}$, reconstructed according to section~\ref{sec:qz}) is also
shown. The number of selected SM background events and the signal efficiency
at this level are presented in Table~\ref{tab:qg3j}. The final event
selection was done by requiring the gluon to have $p_T>75$~GeV$/c$ and
$125<m_{qg}<200$~GeV$/c^2$. No generated QCD ($b\bar{b}$) events passed the
final selection criteria.

\begin{table}[t]
\begin{center}
  \begin{tabular}{|l||r||r|r|r|r|r|}
    \hline
    \multicolumn{1}{|c||}{Description} & \multicolumn{1}{|c||}{Signal}             & \multicolumn{5}{|c|}{Background Processes}\\
    \cline{3-7}
    \multicolumn{1}{|c||}{of}          & \multicolumn{1}{|c||}{$t\rightarrow gq$}  & \multicolumn{1}{|c|}{$Z+$jets} & \multicolumn{1}{|c|}{$Z+W$} & \multicolumn{1}{|c|}{$t\bar t$}  & \multicolumn{1}{|c|}{single $t$} & \multicolumn{1}{|c|}{$W+$jets}\\
    \cline{2-7}
    \multicolumn{1}{|c||}{Cuts}        & \multicolumn{1}{|c||}{$\varepsilon$ (\%)} & \multicolumn{1}{|c|}{Nevt} & \multicolumn{1}{|c|}{Nevt} & \multicolumn{1}{|c|}{Nevt} & \multicolumn{1}{|c|}{Nevt} & \multicolumn{1}{|c|}{Nevt}\\
    \hline
    \hline
    Preselection    & 1.6 & 1356.6 & 427.1 & 24366.7 & 11328.2 & 23320.3 \\
    \hline
    Final Selection & 1.2 &  157.1 &  22.1 &  4985.6 &  1187.9 &  1813.3 \\
    \hline
  \end{tabular}
  \emph{
  \caption{The number of selected background events, normalised to
  $L=10$~fb$^{-1}$, and signal efficiencies
  in the $t\to g q$ channel (``3 jets'') for the
  preselection and final selection levels, obtained with a likelihood-based 
  analysis, are shown.}
  \label{tab:qg3j}}
\end{center}
\end{table}

\begin{figure}
  \psfrag{mat}[rt][rt]{\small $m_{\bar t}$ (GeV/$c^2$)}
  \psfrag{ptat}[rt][rt]{\small ${p_T}_{\bar t}$ (GeV/$c^2$)}

  \includegraphics[width=.48\textwidth]{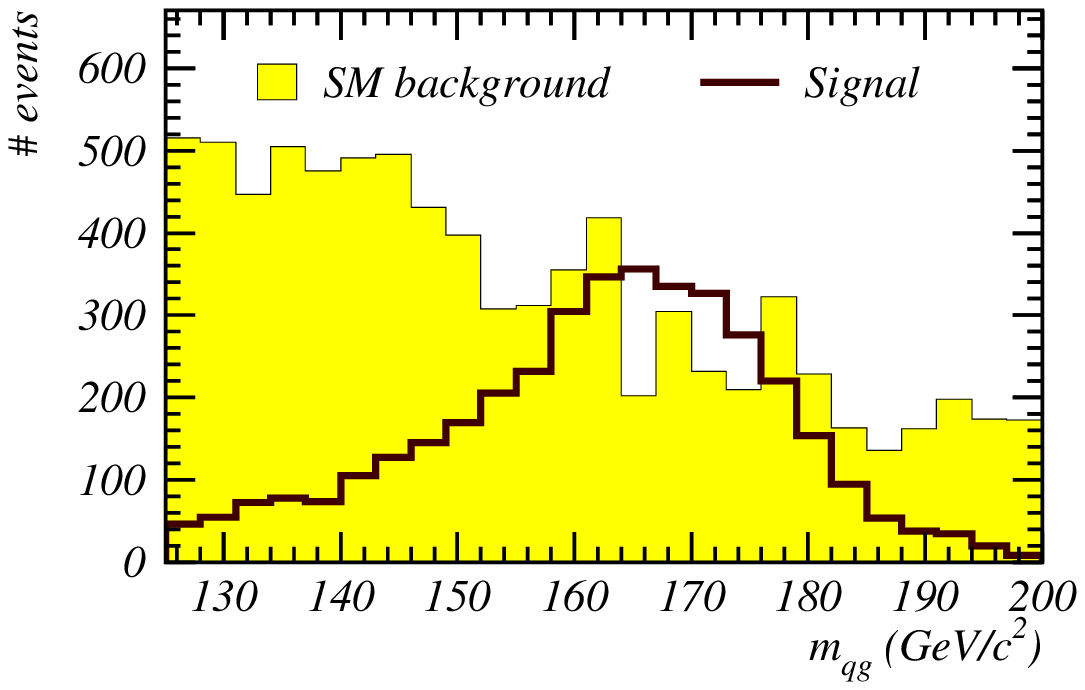}
  \hfill
  \includegraphics[width=.48\textwidth]{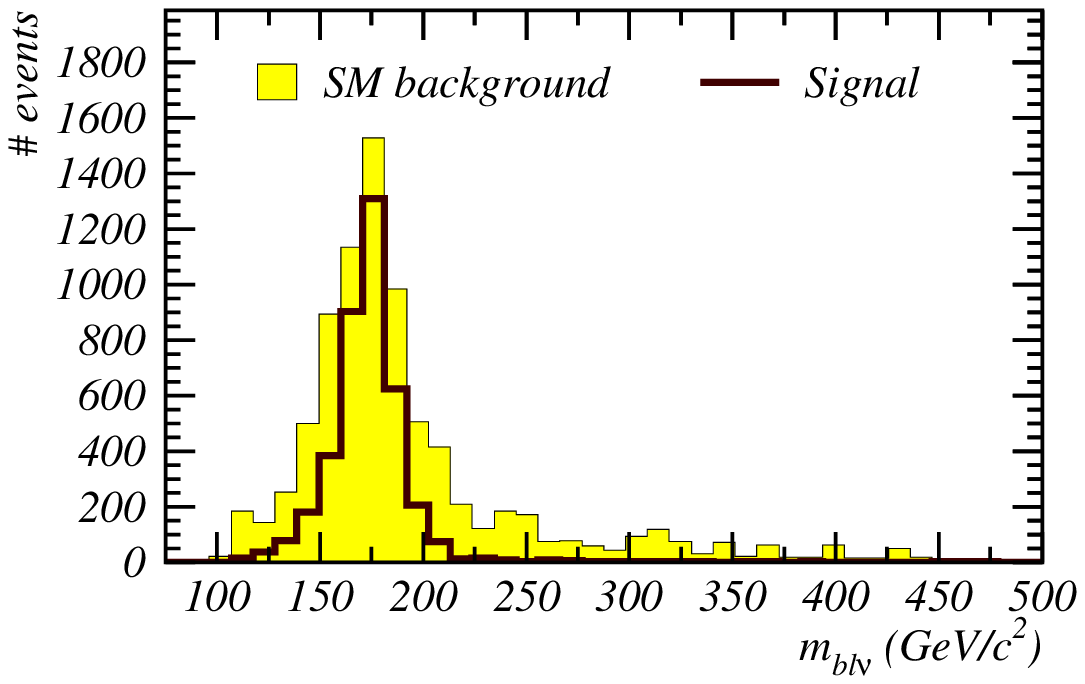}
  \vspace{-1em}

  \hspace{.24\textwidth}a)\hfill b)\hspace{.24\textwidth}\vspace{.5em}

  \includegraphics[width=.48\textwidth]{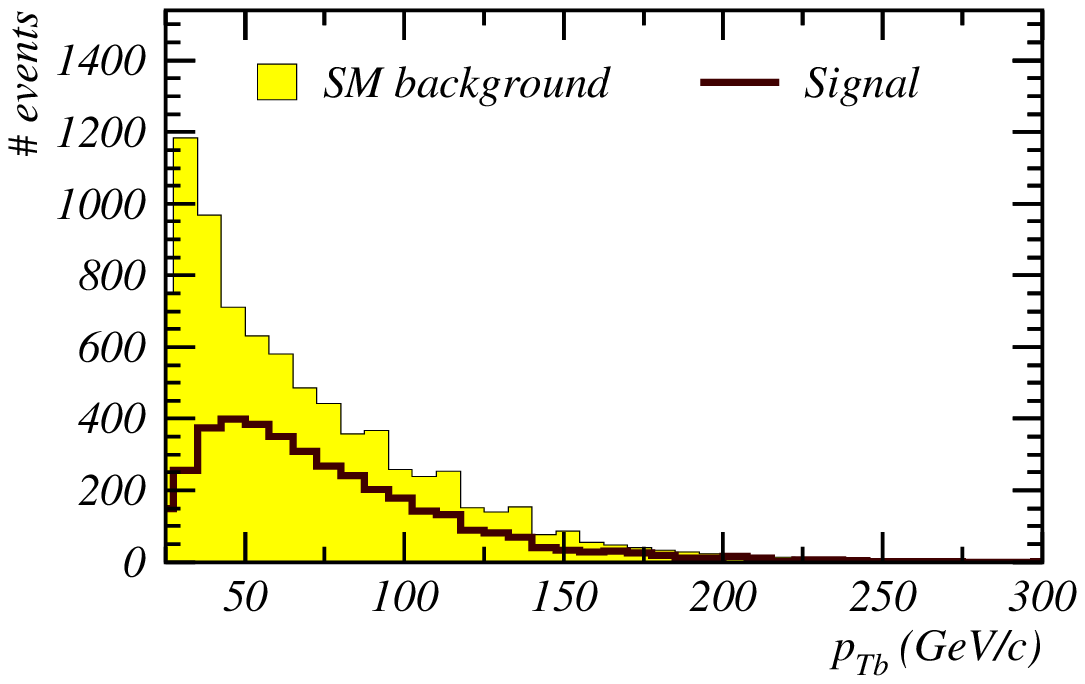}
  \hfill
  \includegraphics[width=.48\textwidth]{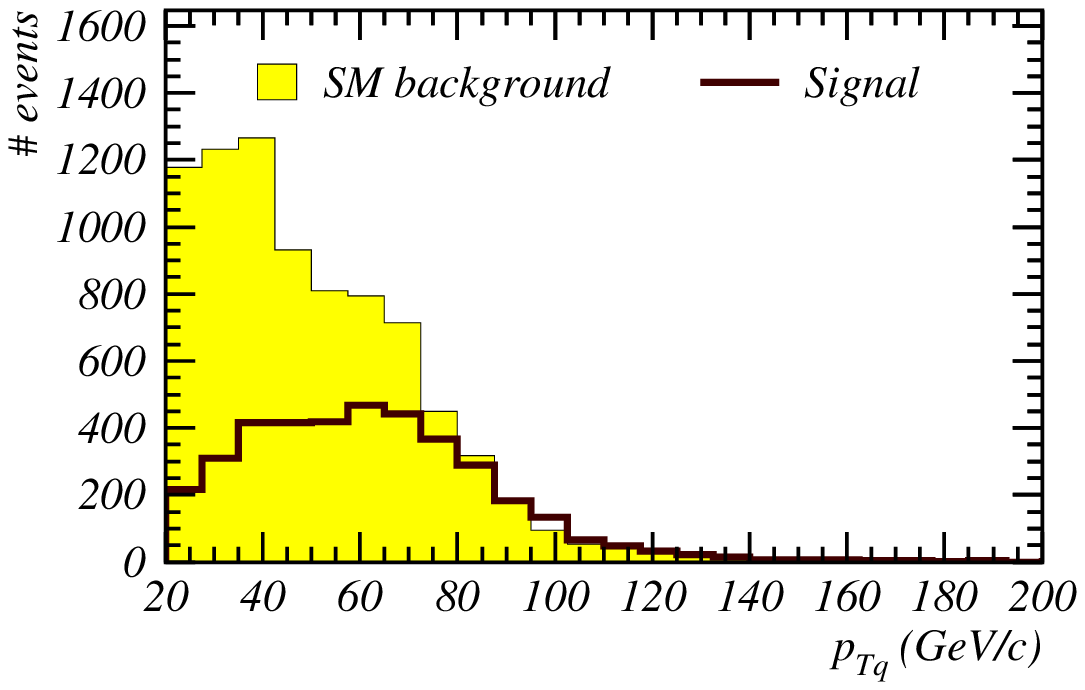}
  \vspace{-1em}

  \hspace{.24\textwidth}c)\hfill d)\hspace{.24\textwidth}\vspace{.5em}

  \includegraphics[width=.995\textwidth]{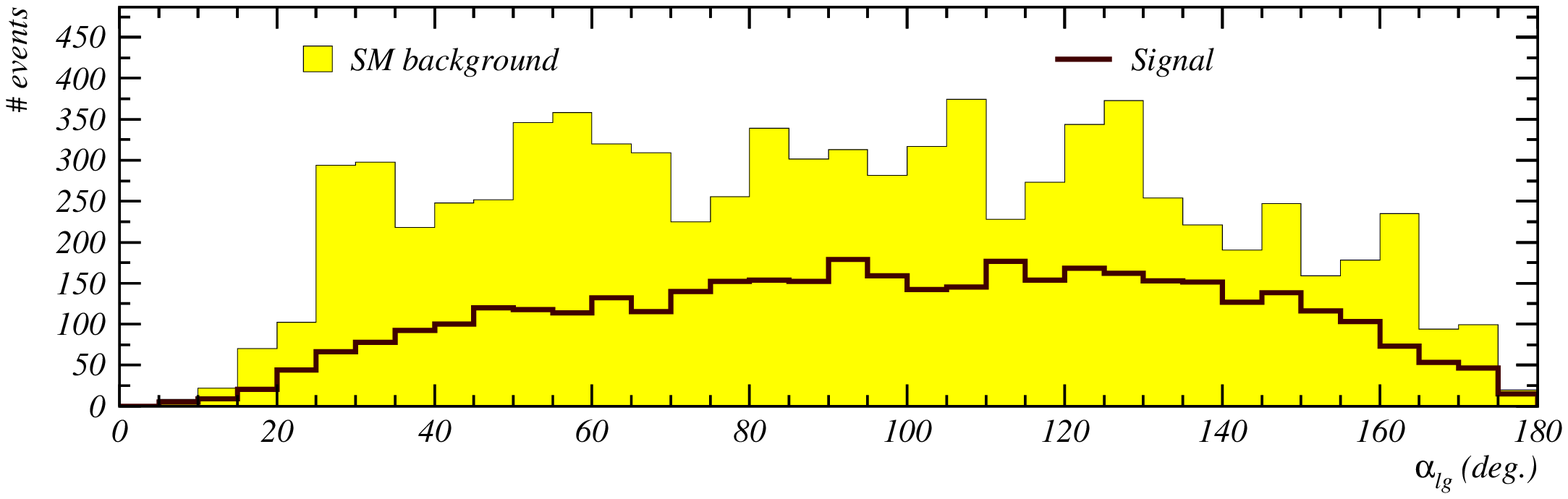}
  \vspace{-1em}

  \hspace{.48\textwidth}e)\hfill

  \vspace*{-1.5em}
  \emph{\caption{The distribution of the variables based on which the
  p.d.f. were built are shown ($t\to g q$ channel --- ``3 jets'')
  a)~the $qg$ invariant mass;
  b)~the $b\ell\nu$ invariant mass;
  c)~transverse momentum of the $b$-jet;
  d)~transverse momentum of the $c(u)$-jet and
  e)~angle between the lepton and the gluon.
  The SM background is normalised to $L=10$~fb$^{-1}$ and the signal has
  an arbitrary normalization, but the same in all plots of this figure.}
  \label{fig:qg3j2}}
\end{figure}

As for the other channels, a likelihood-based type of analysis was adopted, using 
the following variables to build the p.d.f. (\emph{c.f.} Fig.~\ref{fig:qg3j2}):
\begin{itemize}
  \item the $qg$ invariant mass,
  \item the $b\ell\nu$ invariant mass,
  \item transverse momentum of the $b$-jet (${p_T}_{b}$),
  \item transverse momentum of the second non-$b$ jet (${p_T}_{j}$) and
  \item angle between the lepton and the gluon ($\alpha_{\ell g}$).
\end{itemize}

\begin{figure}
  \begin{center}
    \includegraphics[width=.995\textwidth]{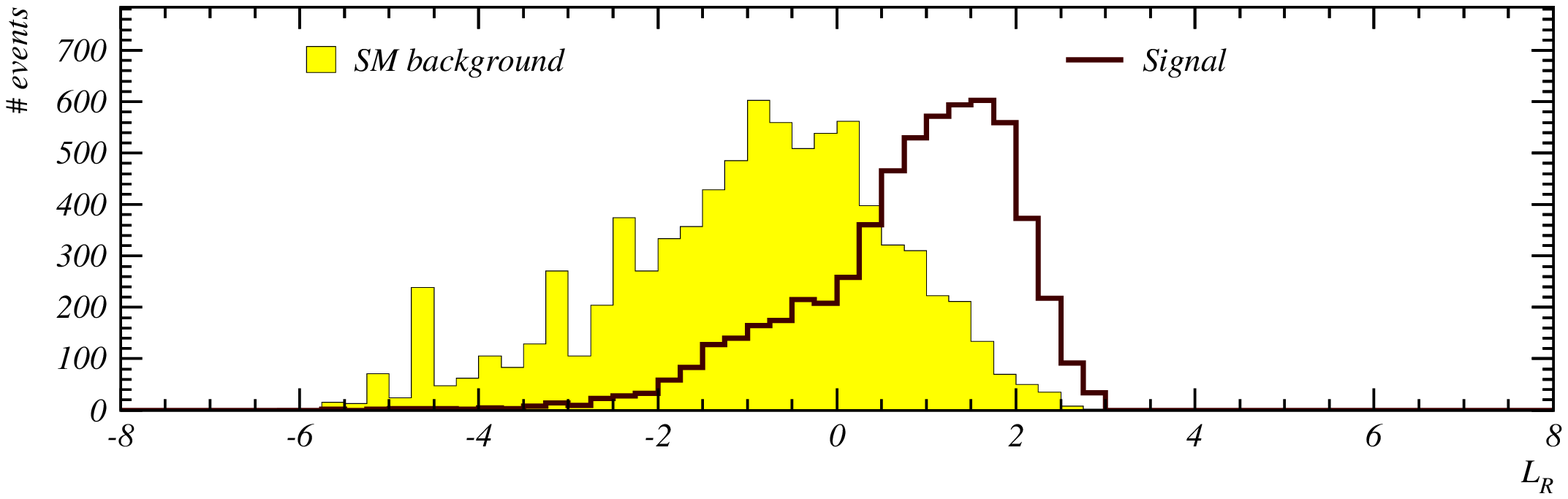}
  \end{center}

  \vspace*{-3em}
  \emph{\caption{Expected background and signal discriminant variable distributions for
  the $t\to g q$ channel with the number of jets equal to three.
  The SM background is normalised to $L=10$~fb$^{-1}$ and the signal has
  an arbitrary normalization.}
  \label{fig:qg3j3}}
\end{figure}

\begin{figure}
  \begin{center}
    \includegraphics[width=.995\textwidth]{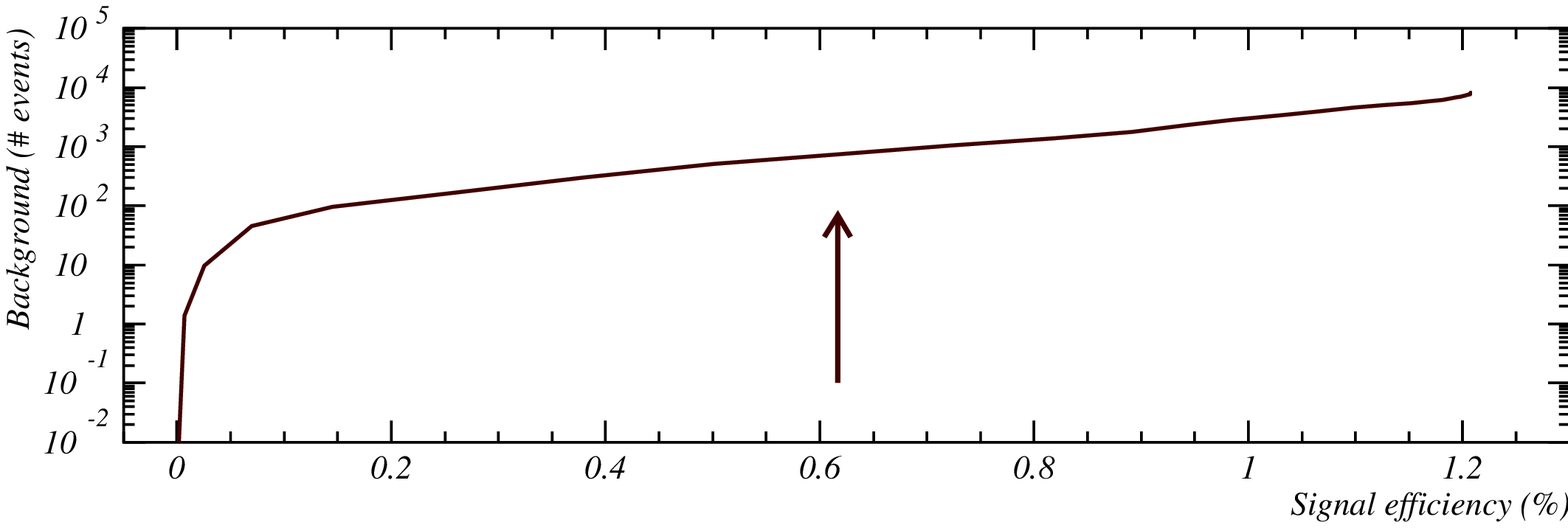}
  \end{center}

  \vspace*{-3em}
  \emph{\caption{The number of expected SM background as a function of 
  the signal efficiency for the $t\to g q$ channel with the number of jets
  equal to three is shown.
  The SM background is normalised to $L=10$~fb$^{-1}$.
  The arrow shows the point with best $S/\sqrt{B}$.}
  \label{fig:qg3j:effvsback}}
\end{figure}

The signal and background discriminant variable distributions are shown in
Fig.~\ref{fig:qg3j3}. The number of expected SM background as a function of
the signal efficiency obtained by cutting the discriminant variable is shown
in Fig.~\ref{fig:qg3j:effvsback}.

\subsubsection{The ``4 jets'' sample}

\begin{figure}
  \psfrag{mat}[rt][rt]{\small $m_{\bar t}$ (GeV/$c^2$)}
  \psfrag{ptat}[rt][rt]{\small ${p_T}_{\bar t}$ (GeV/$c^2$)}

  \includegraphics[width=.48\textwidth]{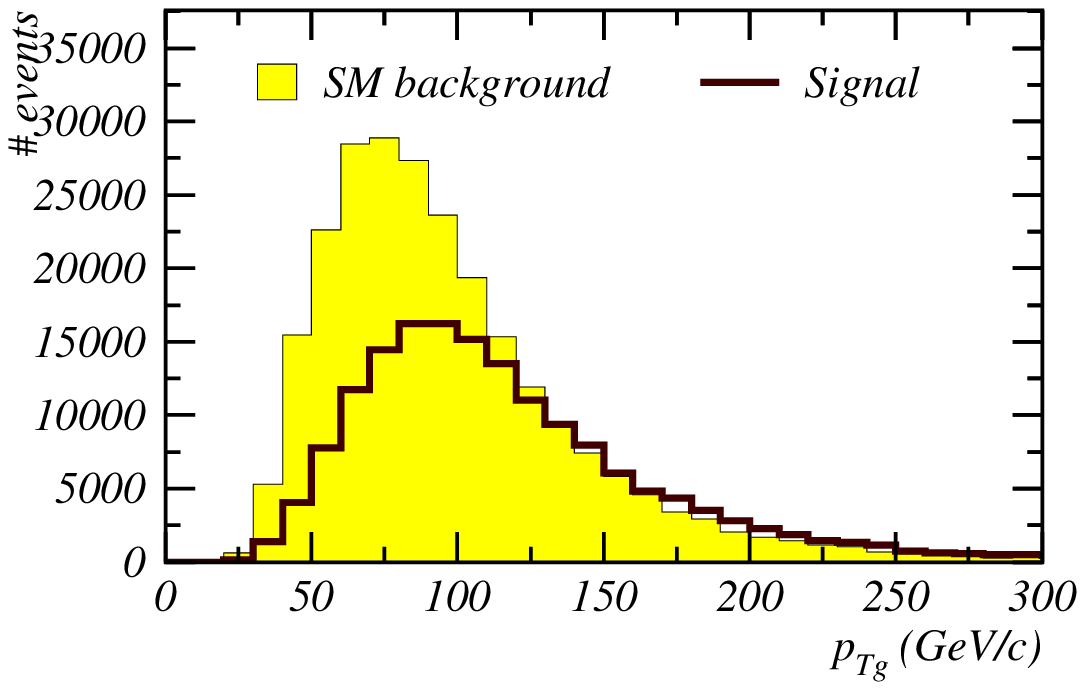}
  \hfill
  \includegraphics[width=.48\textwidth]{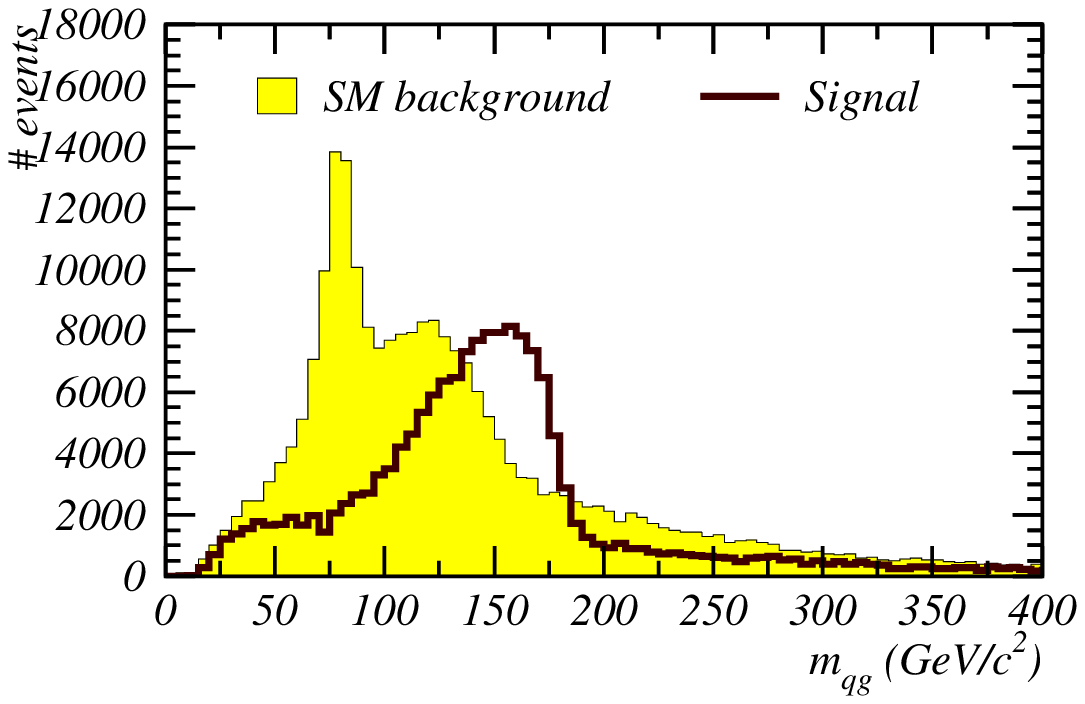}
  \vspace{-1em}

  \hspace{.24\textwidth}a)\hfill b)\hspace{.24\textwidth}\vspace{.5em}

  \includegraphics[width=.998\textwidth]{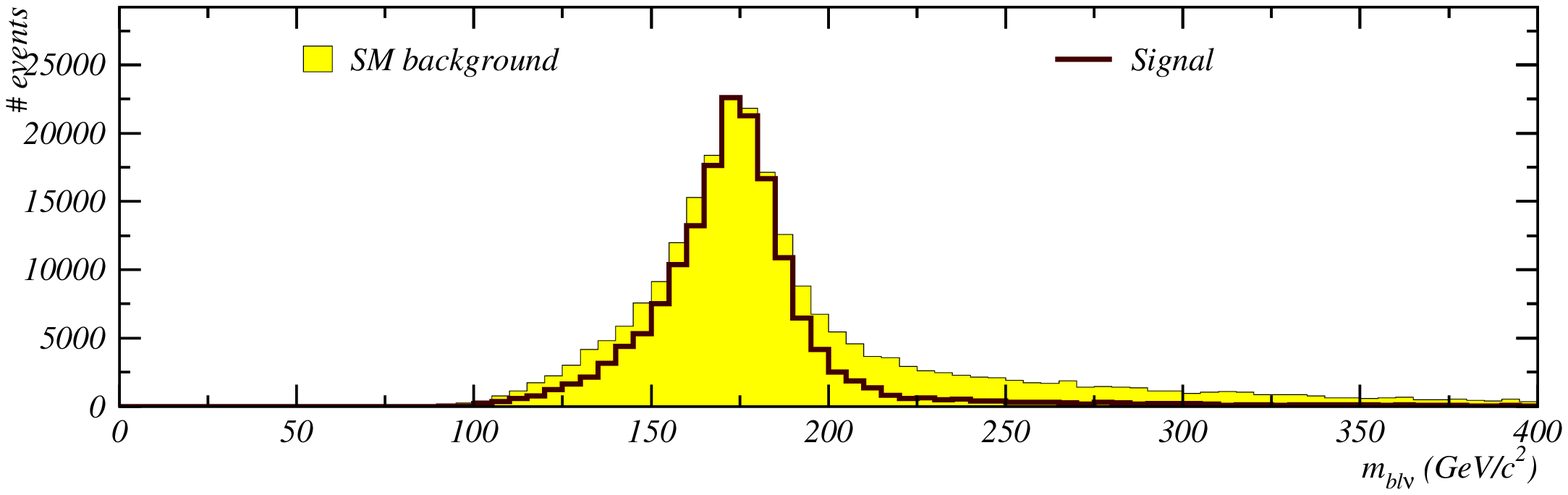}
  \vspace{-1em}

  \hspace{.48\textwidth}c)\hfill

  \vspace*{-2em}
  \emph{\caption{The distributions of relevant variables for the $t\to g
  q$ (``4 jets'') channel are shown after the preselection level:
  a)~transverse momentum of the gluon;
  b)~the $qg$ invariant mass and
  c)~the $b\ell\nu$ invariant mass.
  The SM background is normalised to $L=10$~fb$^{-1}$ and the signal has 
  an arbitrary normalization, but the same in all plots of this figure.}
  \label{fig:qg4j1}}
\end{figure}

For this topology, the preselection was completed by requiring the fourth jet
to have $p_T>20$~GeV$/c$ and $|\eta|<2.5$. Only one $b$-tagged jet, which had
to be among the first four, was allowed in the event. The gluon jet was
assumed to be the non-$b$ jet with the highest transverse momentum. This
distribution is shown in Fig.~\ref{fig:qg4j1}, together with the mass of the
$t$-quark with FCNC decay ($m_{gj}$), reconstructed from the two non-$b$ jets
with the highest transverse momenta. The mass of the $t$-quark with SM decay
is also shown.  The number of selected SM background events and the signal
efficiency at this level are presented in Table~\ref{tab:qg4j}.

 \begin{table}[t]
 \begin{center}
   \begin{tabular}{|l||r||r|r|r|r|r|}
     \hline
     \multicolumn{1}{|c||}{Description} & \multicolumn{1}{|c||}{Signal}             & \multicolumn{5}{|c|}{Background Processes}\\
     \cline{3-7}
     \multicolumn{1}{|c||}{of}          & \multicolumn{1}{|c||}{$t\rightarrow gq$}  & \multicolumn{1}{|c|}{$Z+$jets} & \multicolumn{1}{|c|}{$Z+W$} & \multicolumn{1}{|c|}{$t\bar t$}  & \multicolumn{1}{|c|}{single $t$} & \multicolumn{1}{|c|}{$W+$jets}\\
     \cline{2-7}
     \multicolumn{1}{|c||}{Cuts}        & \multicolumn{1}{|c||}{$\varepsilon$ (\%)} & \multicolumn{1}{|c|}{Nevt} & \multicolumn{1}{|c|}{Nevt} & \multicolumn{1}{|c|}{Nevt} & \multicolumn{1}{|c|}{Nevt} & \multicolumn{1}{|c|}{Nevt}\\
     \hline
     \hline
     Preselection    & 5.7 & 1171.0 & 305.2 & 216679.9 & 14263.1 & 12651.2 \\
     \hline
     Final Selection & 1.2 &   64.3 &   7.1 &   9142.1 &   453.3 &   379.5 \\
     \hline
   \end{tabular}
  \emph{\caption{The number of selected background events, normalised to
    $L=10$~fb$^{-1}$, and signal efficiencies
    in the $t\to g q$ channel (``4 jets'') for the
    preselection and final selection levels, obtained with a 
    likelihood-based analysis, are shown.}
  \label{tab:qg4j}}
 \end{center}
 \end{table}

The final selection was defined by requiring the gluon transverse momentum to
be above $100$~GeV$/c$ and the reconstructed mass of the $t$-quark with FCNC
decay above $150$~GeV$/c^2$ and below $190$~GeV$/c^2$. As for the ``3 jets''
channel, no generated QCD ($b\bar{b}$) passed the final selection criteria.

\begin{figure}
  \psfrag{mat}[rt][rt]{\small $m_{\bar t}$ (GeV/$c^2$)}
  \psfrag{ptat}[rt][rt]{\small ${p_T}_{\bar t}$ (GeV/$c^2$)}

  \includegraphics[width=.48\textwidth]{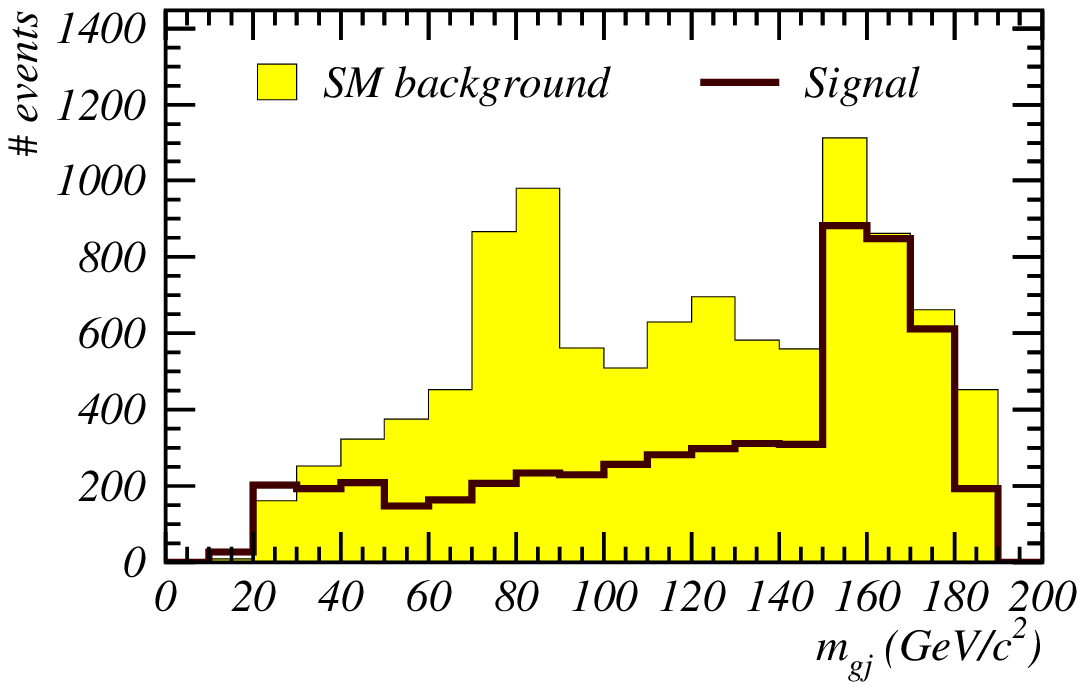}
  \hfill
  \includegraphics[width=.48\textwidth]{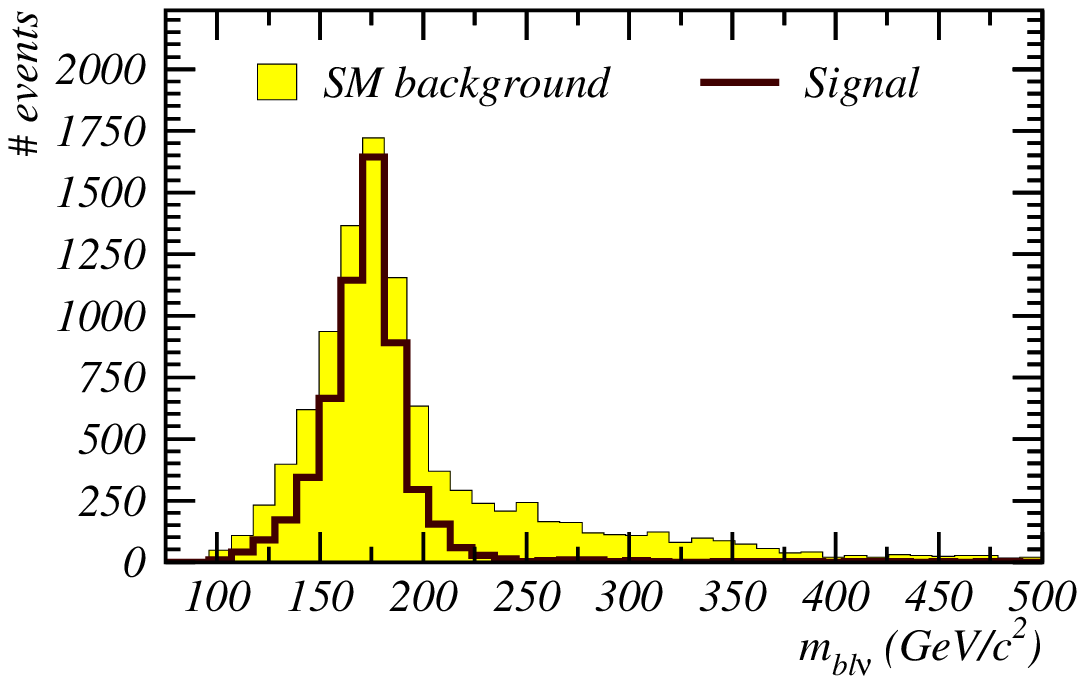}

  \hspace{.24\textwidth}a)\hfill b)\hspace{.24\textwidth}\vspace{.5em}

  \includegraphics[width=.48\textwidth]{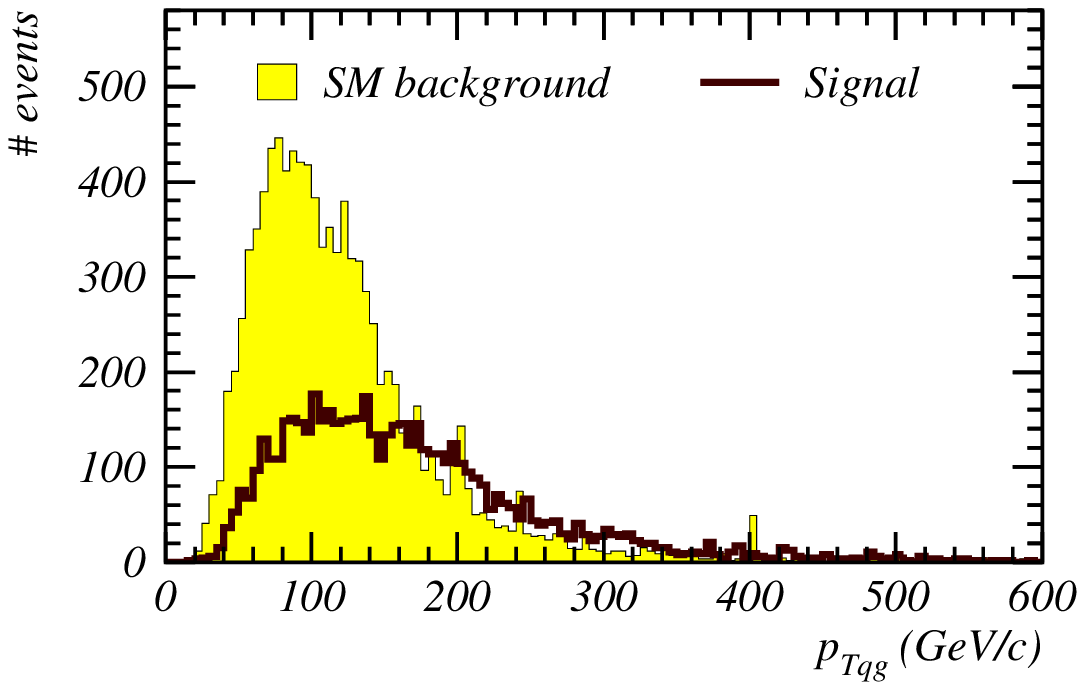}
  \hfill
  \includegraphics[width=.48\textwidth]{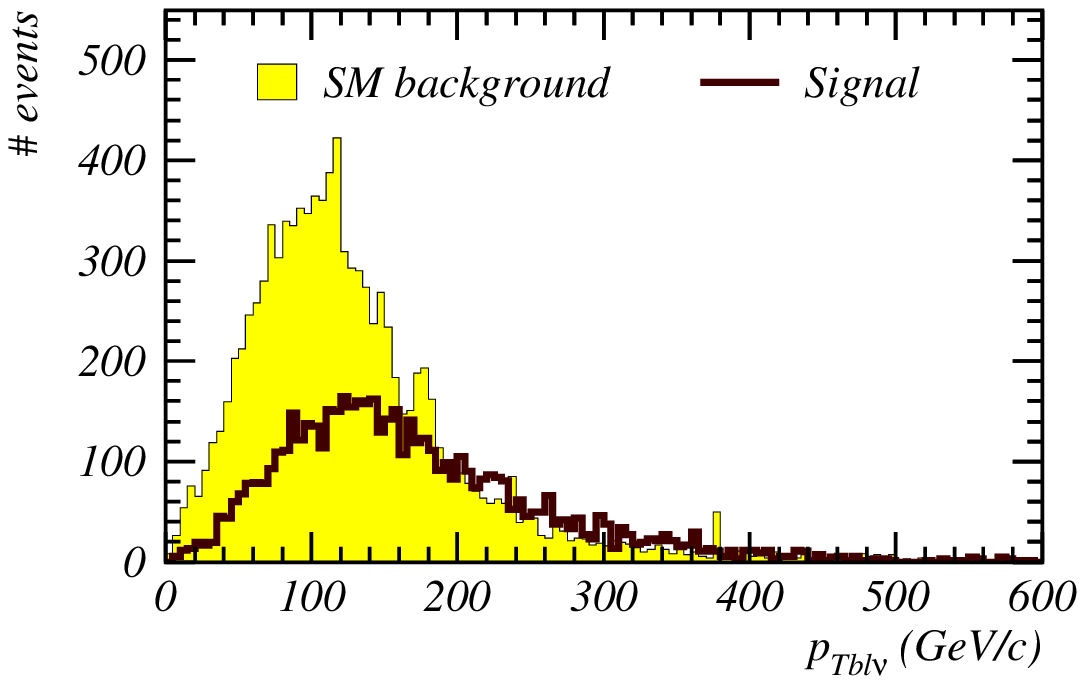}  
  \vspace{-1em}

  \hspace{.24\textwidth}c)\hfill d)\hspace{.24\textwidth}\vspace{.5em}

  \includegraphics[width=.48\textwidth]{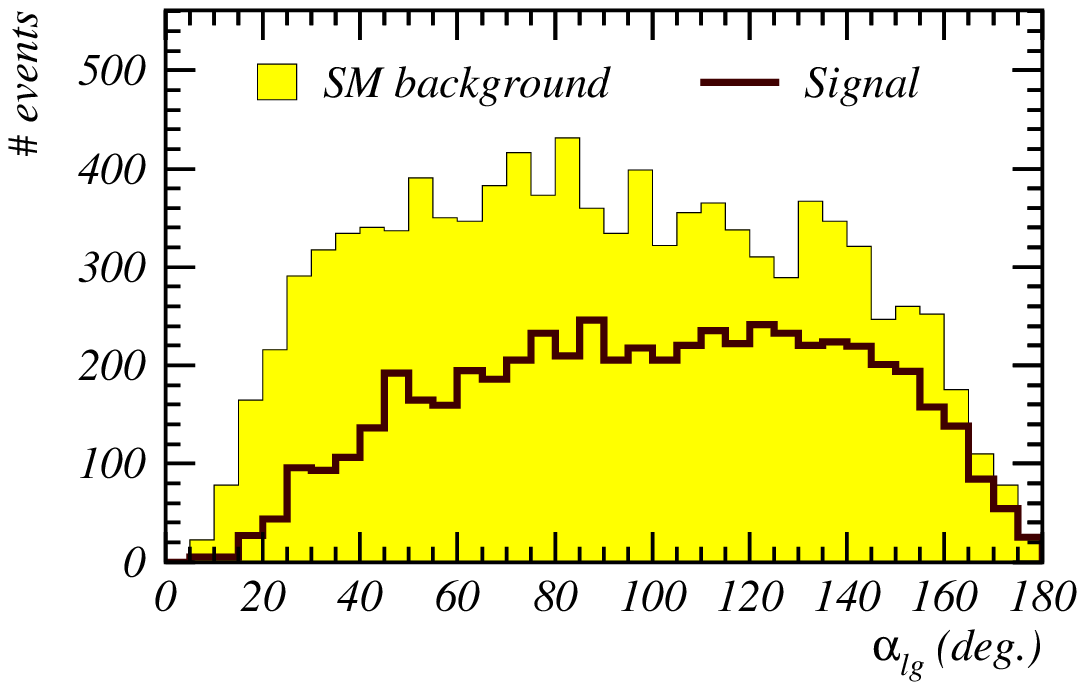}
  \hfill
  \includegraphics[width=.48\textwidth]{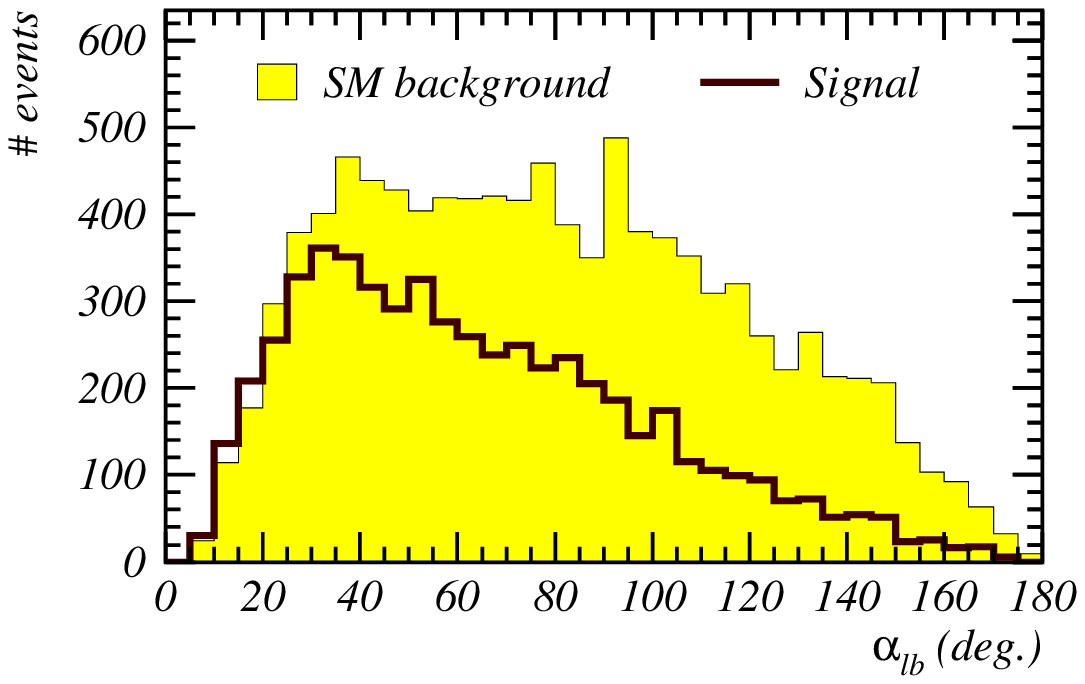}
  \vspace{-1em}
  
  \hspace{.24\textwidth}e)\hfill f)\hspace{.24\textwidth}\vspace{.5em}

  \includegraphics[width=.995\textwidth]{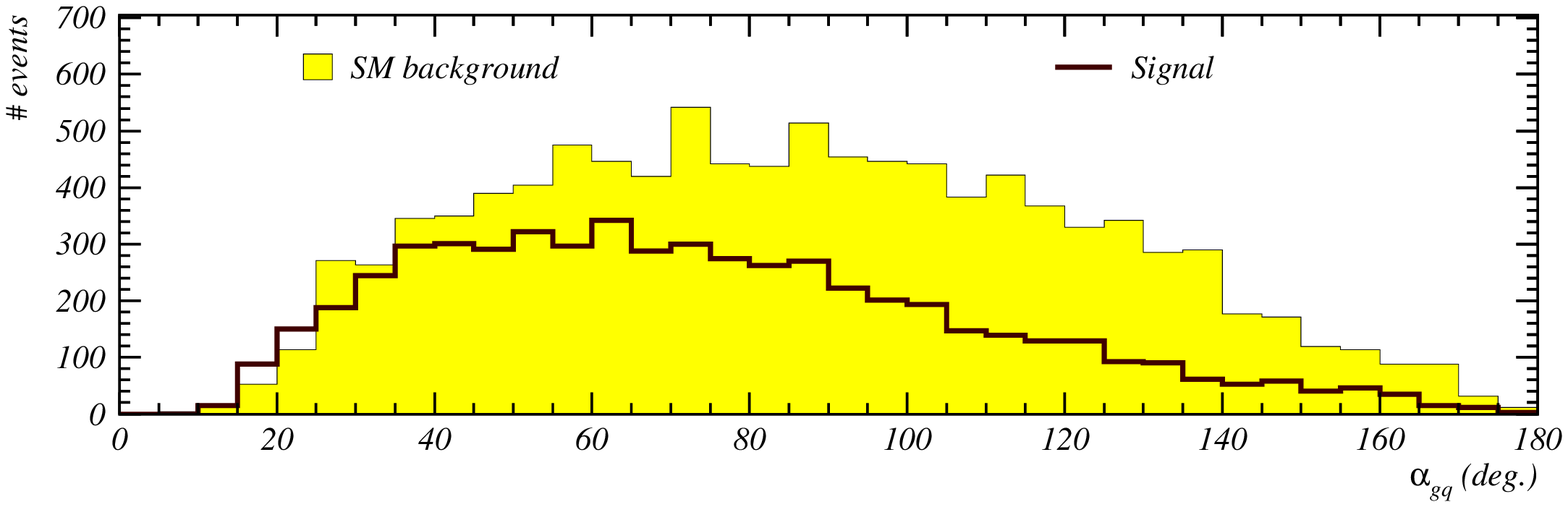}
  \vspace{-1em}

  \hspace{.48\textwidth}g)\hfill
  
  \vspace*{-1.5em}
  \emph{\caption{The distribution of the variables based on which the
  p.d.f. were built are shown ($t\to g q$ channel --- ``4 jets'')
  a)~minimum invariant mass of the first and the second non-$b$ jets or the
     first and the third non-$b$ jets;
  b)~the $b\ell\nu$ invariant mass;
  c)~reconstructed transverse momentum of the $qg$;
  d)~reconstructed transverse momentum of the $b\ell\nu$;  
  e)~angle between the lepton and the gluon;
  f)~angle between the lepton and the $b$-jet and
  g)~angle between the gluon and the second non-$b$ jet.
  The SM background is normalised to $L=10$~fb$^{-1}$ and the signal has
  an arbitrary normalization, but the same in all plots of this figure.}
  \label{fig:qg4j2}}
\end{figure}

After the final selection, the p.d.f. were built based on the following 
physical distributions (\emph{c.f.} Fig.~\ref{fig:qg4j2}):
\begin{itemize}
  \item minimum invariant mass of the leading and the second non-$b$
  jets or the leading and the third non-$b$ jets ($m_{gj}$),
  \item the $b\ell\nu$ invariant mass,
  \item transverse momentum of $qg$,
  \item transverse momentum of $b\ell\nu$,
  \item angle between the lepton and the gluon ($\alpha_{\ell g}$),
  \item angle between the lepton and the $b$-jet ($\alpha_{\ell b}$) and
  \item angle between the gluon and the second non-$b$ jet ($\alpha_{gq}$).
\end{itemize}

\begin{figure}
  \begin{center}
    \includegraphics[width=.995\textwidth]{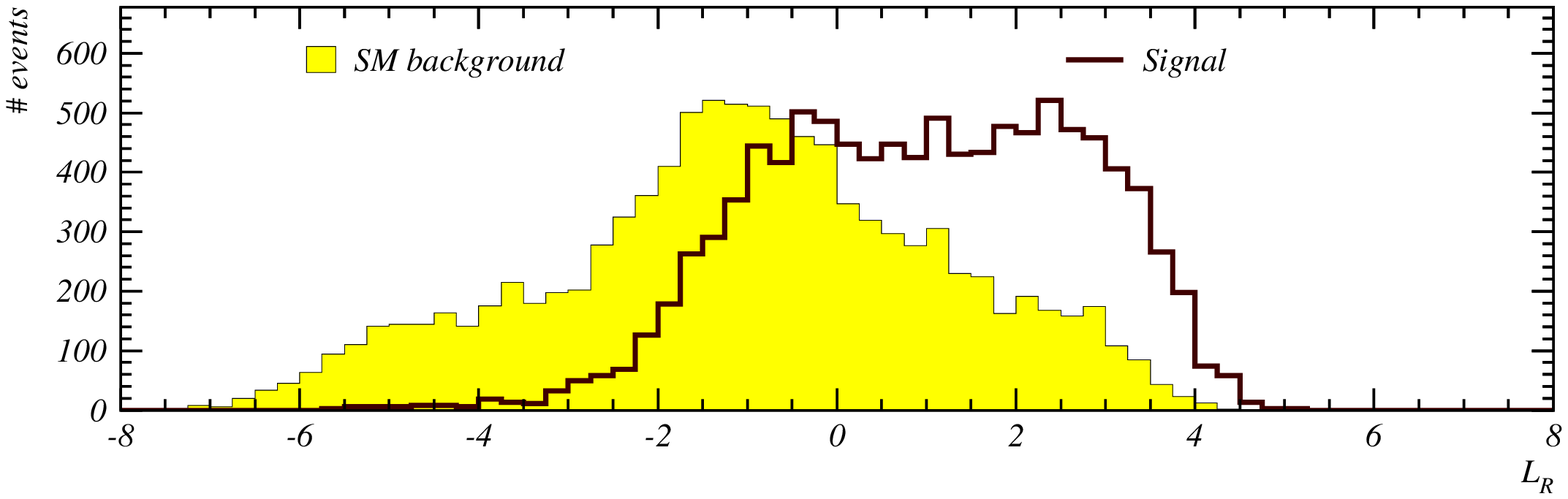}
  \end{center}

  \vspace*{-3em}
  \emph{\caption{Expected background and signal discriminant variable 
  distributions for the $t\to g q$ channel with the number of jets 
  greater than three. The SM background is normalised to 
  $L=10$~fb$^{-1}$ and the signal has an arbitrary normalization.}
  \label{fig:qg4j3}}
\end{figure}

\begin{figure}
  \begin{center}
    \includegraphics[width=.995\textwidth]{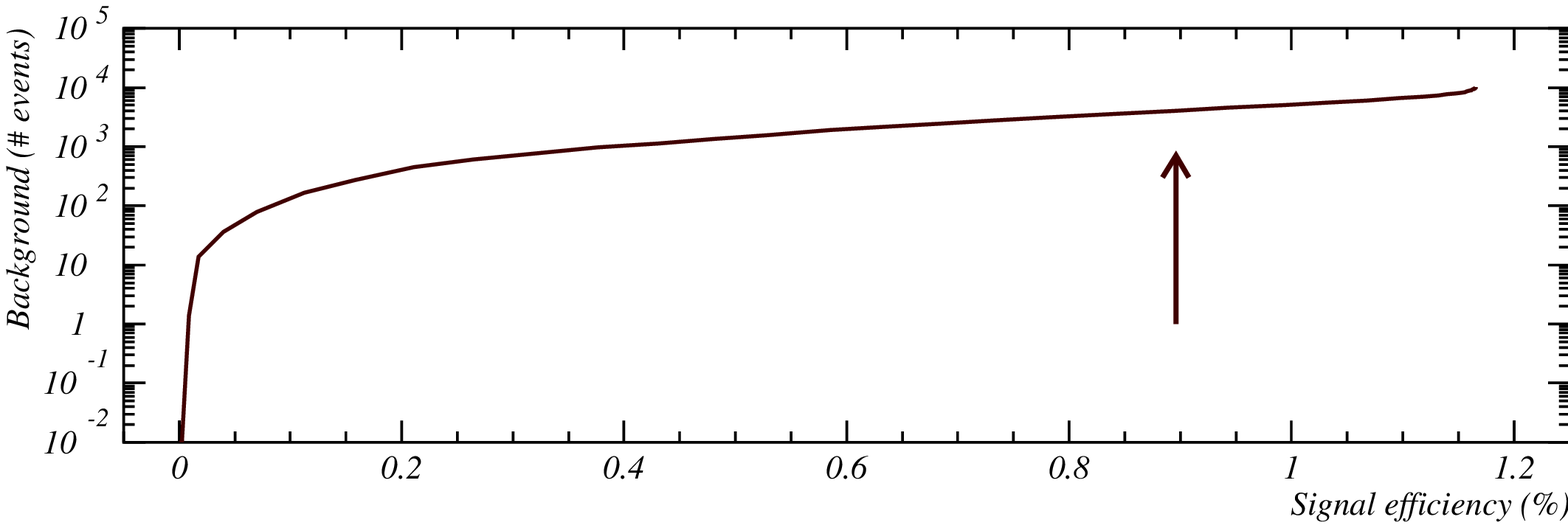}
  \end{center}

  \vspace*{-3em}
  \emph{\caption{The number of expected SM background as a function of 
  the signal efficiency for the $t\to g q$ channel with the number of jets
  greater than three is shown.
  The SM background is normalised to $L=10$~fb$^{-1}$.
  The arrow shows the point with best $S/\sqrt{B}$.}
  \label{fig:qg4j:effvsback}}
\end{figure}

The discriminant variable distributions for signal and SM expectation are
shown in Fig.~\ref{fig:qg4j3}, while the number of expected SM background as
a function of the signal efficiency obtained by cutting the discriminant
variable is shown in Fig.~\ref{fig:qg4j:effvsback}.
 
\section{Results and systematic studies \label{results}}

Expected top quark FCNC decay branching ratios sensitivities of the
ATLAS experiment were estimated for both the cut-based and
likelihood-based analysis under two different hypothesis, as explained in
the next subsections.

\subsection{Branching ratio sensitivity ($\boldsymbol{5\sigma}$ significance 
discovery hypothesis) }

Assuming a signal discovery with a $5\sigma$ significance, the branching
ratio ($BR$) sensitivity for each channel studied is estimated by:
\begin{equation} BR = \frac{5\sqrt{B\times \varepsilon_\ell}}{2\times L
\times \sigma(t\bar t_{SM}) \times \varepsilon_t \times \varepsilon_\ell} \,
, \end{equation} where $\sigma(t\bar t_{SM})=833$~pb~\cite{ttcs} is the NLO
calculation of the SM cross-section for $t\bar t$ production in $pp$
collisions at $\sqrt{s}=14$~TeV. $B$ is the total number of selected
background events, $\varepsilon_t$ is the signal efficiency convoluted with
the appropriate branching ratios and $\varepsilon_\ell=0.9^n$ is the charged
leptons identification efficiency ($n$ is the number of leptons required for
each channel). The factor 2 in the denominator takes into account the $t$ and
$\bar t$ contributions to the $BR$.

To evaluate the expected branching ratio sensitivities for a $5\sigma$
signal significance of discovery in the cut-based analyses, the
kinematic cuts were applied in sequence for the signal and backgrounds.
In the channels studied using likelihood-based analyses, the expected
branching ratio sensitivities were evaluated after applying cuts to the
discriminant variables, as given in Table~\ref{tab:5sigmacut} (see also 
Fig.~\ref{fig:qz:effvsback}, \ref{fig:qgamma:effvsback}, 
\ref{fig:qg3j:effvsback} and \ref{fig:qg4j:effvsback}). These
cuts were optimised according to the best $S/\sqrt{B}$ ($S$ is the number of
selected signal events). The expected branching ratio sensitivities for a
$5\sigma$ discovery are shown in Table~\ref{tab:5sigma}.

\begin{table}[t]
\begin{center}
  \begin{tabular}{|l|l||r|r|r|}
    \hline
    \multicolumn{2}{|l||}{channel}                     & $L_R$ cut & $B$    & $\varepsilon_t$ (\% ) \\
    \hline
    \hline
    \multicolumn{2}{|l||}{$t\to Z q$}               & $>5.62$   & 0.50   & 0.06                  \\
    \hline
    \multicolumn{2}{|l||}{$t\to \gamma q$}          & $>2.71$   & 3.48   & 0.62                  \\
    \hline
    \multirow{2}{*}{$t\to g q$}        & ``3 jets'' & $>1.13$   & 734.1  & 0.20                  \\
    \cline{2-5}
                                          & ``4 jets'' & $>-0.38$  & 4033.9 & 0.29                  \\
    \hline
  \end{tabular}
  \emph{\caption{The number of selected background events (normalised to 
  $L=10$~fb$^{-1}$) and signal efficiencies (convoluted with the 
  appropriate branching ratios) for each channel of the likelihood-based analyses 
  after the specified $L_R$ cut are shown.}
  \label{tab:5sigmacut}}
\end{center}
\end{table}

\begin{table}[t]
\begin{center}
  \begin{tabular}{|l|l|l|r|r|}
  \hline
  channel                            & type                              &             & $BR$ ($L=10$~fb$^{-1}$)         & $BR$ ($L=100$~fb$^{-1}$) \\
  \hline
  \hline
  \multirow{3}{*}{$t\to Z q$}     & \multirow{2}{*}{cut-based}        & hadronic    & $1.7\times 10^{-3}$\cite{dodd}  & $5.0\times 10^{-4}$ \\
  \cline{3-5}
                                     &                                   & leptonic    & $4.7\times 10^{-4}$\cite{dodd}  & $1.1\times 10^{-4}$ \\
  \cline{2-5}
                                     & likelihood-based                  & leptonic    & $4.4\times 10^{-4}$             & $1.4\times 10^{-4}$ \\
  \hline
  \hline
  \multirow{2}{*}{$t\to\gamma q$} & cut-based                         &             & --                              & $1.0\times 10^{-4}$\cite{dodd}      \\
  \cline{2-5}
                                     & likelihood-based                  &             & $9.4\times 10^{-5}$             & $3.0\times 10^{-5}$  \\
  \hline
  \hline
  \multirow{2}{*}{$t\to g q$}     & \multirow{2}{*}{likelihood-based} & ``3 jets''  & $4.3\times 10^{-3}$             & $1.4\times 10^{-3}$  \\
  \cline{3-5}
                                     &                                   & ``4 jets''  & $6.9\times 10^{-3}$             & $2.2\times 10^{-3}$  \\
  \hline
  \end{tabular}
  \emph{\caption{The branching ratio sensitivity for each channel in the
  $5\sigma$ discovery hypothesis is shown. The results for a luminosity of
  $L=10$ and $100$~fb$^{-1}$ are presented. The values shown for the 
  likelihood-based analyses were obtained after applying the cuts described 
  in table~\ref{tab:5sigmacut}. The values presented for the cut-based 
  analyses of the $t\to Zq$ channel, with $L=10$~fb$^{-1}$, and of 
  the $t\to \gamma q$ channel were taken from Ref.~\cite{dodd}
}
  \label{tab:5sigma}}
\end{center}
\end{table}

\subsection{95\% confidence level limits (hypothesis of absence of signal)}

In the absence of a FCNC top decay signal, expected limits at 95\% CL can be
derived. These limits were obtained for both the cut-based and the
likelihood-based analyses, setting the charged lepton identification
efficiency to $90\%$.

For the cut-based analyses of the $t\to Zq$ channel, the 95\% CL
upper limits were evaluated considering an integrated luminosity of
100~fb$^{-1}$. Assuming the Poisson processes with backgrounds, 95\% CL
upper limits on the number of signal events for both decay modes were
derived. The modified frequentist likelihood method~\cite{ar} was used
to evaluate the 95\% CL upper limits for the likelihood-based analyses. The
full information of the discriminant variables were used to derive 95\%
CL upper limits on the number of signal events for each channel. No cuts
on the discriminant variables were used. Using the NLO calculation for
$\sigma(t\bar t_{SM})$, these limits were then converted into limits on
the branching ratio for each of the studied FCNC top decay channels. The
expected 95\% confidence level limits on the branching ratios are
summarised in Table~\ref{tab:lim95}.

\begin{table}[b]
\begin{center}
  \begin{tabular}{|l|l|l|r|r|}
  \hline
  channel                            & type                              &                        & $BR$ ($L=10$~fb$^{-1}$) & $BR$ ($L=100$~fb$^{-1}$) \\
  \hline
  \hline
  \multirow{4}{*}{$t\to Z q$}     & \multirow{3}{*}{cut-based}        & hadronic               & --                      & $2.7\times 10^{-4}$ \\
  \cline{3-5}
                                     &                                   & leptonic               & --                      & $6.3\times 10^{-5}$ \\
  \cline{3-5}
                                     &                                   & combined               & --                      & $5.5\times 10^{-5}$      \\
  \cline{2-5}
                                     & likelihood-based                  & leptonic               & $3.1\times 10^{-4}$     & $6.1\times 10^{-5}$      \\
  \hline
  \hline
  $t\to\gamma q$                  & likelihood-based                  &                        & $4.1\times 10^{-5}$     & $1.2\times 10^{-5}$      \\
  \hline
  \hline
  \multirow{3}{*}{$t\to g q$}     & \multirow{3}{*}{likelihood-based} & ``3 jets''             & $1.6\times 10^{-3}$     & $4.8\times 10^{-4}$      \\
  \cline{3-5}
                                     &                                   & ``4 jets''             & $2.4\times 10^{-3}$     & $7.5\times 10^{-4}$      \\
  \cline{3-5}
                                     &                                   & combined               & $1.3\times 10^{-3}$     & $4.2\times 10^{-4}$      \\
  \hline
  \end{tabular}
  \emph{\caption{The expected 95$\%$ confidence level limits on the 
  FCNC top decays branching ratio in the absence of signal hypothesis 
  are shown. The results for a luminosity of $L=10$ and 100~fb$^{-1}$ 
  are presented.}
  \label{tab:lim95}}
\end{center}
\end{table}

\subsection{Systematic uncertainties and analyses stability\label{systematics}}

The effect of different systematic sources of uncertainty on the limits
evaluation was studied for both the cut-based and the likelihood-based 
analyses. This estimation was done by considering the changes on the 
central values of the signal efficiency, number of background events and 
likelihood ratio distributions.

For the cut-based analysis of the $t\to Z q$ channel several systematic
uncertainties were studied. The effect of the mass window cut applied to the
top quark (which decayed through the FCNC channel) was studied by selecting
events in a more restrictive window i.e., $m_t \pm 12$~GeV$/c^2$
($\sim\sigma$). This results in a increase $12\%$ ($8\%$) on the 95\% CL
limit for the hadronic (leptonic) mode. Varying the $p_T$ cut applied to the
jets from 50 to 40~GeV$/c^2$ (leptonic mode) results in a relative change of
the limit of $23\%$~\cite{ld}. This variation significantly increases the $t\bar 
t$, the $WZ$ and partly the $Z+$jets backgrounds. Changing the lepton
isolation criteria (to $\Delta R=0.2$) gives a relative change of $1\%$ in
the 95\% CL limit.

For the likelihood-based analysis of the leptonic mode of $t\to Z q$, the
$t\to\gamma q$ and the $t\to g q$ channels, the following systematic
uncertainties were considered. The effect of the top mass uncertainty was
evaluated using different Monte Carlo samples with $m_t=170$~GeV$/c^2$ and
$m_t=180$~GeV$/c^2$. This systematic uncertainty affects both the event
kinematics (and consequently the discriminant variables shape) and the value
of the $t\bar t$ cross-section (used in the limits evaluation). 
The overall theoretical uncertainty on $\sigma (t\bar t)$ was estimated in 
reference~\cite{yr}. The effect of this uncertainty was studied by allowing a 
change of 12\% on the central value of $\sigma (t\bar t)$, cross-section used 
both in the $t\bar t_{SM}$ background normalisation and in the $BR$ limits 
evaluation, assuming a negligible error on the measurement itself. If the 
error on the cross-section measurement is, for instance, 5\%, the ATLAS 
sensitivity will be degraded but the change will not affect the order of 
magnitude of the results shown in this paper.
For the $t\to Z q$ and the $t\to\gamma q$ channels, a 5\% error gives a 
maximum change on the limit of 5\%. For the $t\to gq$ channels, where the 
expected number of background is more important, the limit can change by a 
factor 2 to 3 (depending if it is the 3 or 4 jets topology). A precise 
measurement of the $t\bar t$ cross-section is, for this reason, of utmost 
importance.
The CTEQ 5L PDF set was used in the Monte Carlo
generation. A different PDF set (CTEQ 4M~\cite{pythia,toprex}) was used to
estimate the effect of this choice on the event kinematics. As mentioned in
section~\ref{simulation}, the ATLFASTB package was used to simulate the
$b$-tag algorithm with a $b$-tag efficiency of 60\%. In order to study the
impact of the algorithm with a different efficiency, the $b$-tagging
efficiencies of 50\%{} and 70\%{} were also considered. This source of
uncertainty affects the signal efficiency, background estimation and
discriminant variable shapes. The impact of the knowledge of the absolute jet
energy scale was estimated by recalibrating the reconstructed jet energy. A
miscalibration of $\pm 1$\% for light jets and $\pm 3$\% for $b$-jets was
used. This uncertainty was found to have a negligible effect on the signal
efficiency, background estimation and discriminant variable shapes.
A jet energy miscalibration of $\pm 5$\% for all jets was
also considered. For the $t\to Z q$, $t\to \gamma q$ and $t\to g q$
(``3 jets'') channels the relative changes on the 95\% CL expected limits were
found to be below 7\%. For the most difficult channel ($t\to g q$ -- ``4
jets'') this effect is more important (up to 12\%), due to the tighter
selection criteria used to reject the large contribution from background. The
stability of the cut-based analysis was studied by changing the preselection
and final selection (typically a $\pm$10\% variation on the cut values was
considered). The discriminant variables were computed using the probability
density function sets described in section~\ref{analyses}. In order to
estimate the effect of a different p.d.f. set, the following changes were
studied: in the $t\to Z q$ channel, the $\bar t$ reconstruction was done
by considering the jet closest to the reconstructed $Z$ in the invariant mass
evaluation. Similarly, the $\bar t$ mass reconstruction in the $t\to
\gamma q$ channel was done using the jet closest to the leading $\gamma$.
Moreover, the $t$ mass was included in the p.d.f. set and the multiplicity of
jets with $|\eta|<2.5$ was chosen as p.d.f. (instead of the jet
multiplicity). In the $t\to g q$ channel, $\Delta R$ was used instead of
the angles, in the p.d.f.s definition.

The absolute value of the maximum relative effect on the 95\% confidence
level expected limits on each considered source of systematic uncertainty
(the reference values are those presented in Table~\ref{tab:lim95}) is shown
in Table~\ref{tab:syst} ($L=10$~fb$^{-1}$). Although differences up to $20$\%
were observed (caused by the uncertainty of the top mass), the order of
magnitude of the expected limits on the $BR$ is not affected by any of the
systematic uncertainties considered. Moreover, the change on the selection
criteria and on the p.d.f. sets do not have a significant impact on the
results.

\begin{table}[bt]
\begin{center}
\begin{tabular}{|c||c||c||c|c|}
  \hline
  \multirow{2}{*}{Source} & \multirow{2}{*}{$t\to Z q$} & \multirow{2}{*}{$t\to \gamma q$} & \multicolumn{2}{|c|}{$t\to g q$}  \\
  \cline{4-5}
                                &             &              & (3 jets)    & (4 jets)   \\
  \hline
  \hline
  top mass                      &  18\%       &  13\%        &   8\%       &  7\%        \\
  \hline
  $\sigma (t\bar t)$            &  11\%       &  11\%        &   9\%       &  7\%        \\
  \hline
  PDFs choice                   &  15\%       &   7\%        &   3\%       &  6\%        \\
  \hline
  $b$-tag algorithm efficiency  &  16\%       &   5\%        &  18\%       & 17\%        \\
  \hline
  jet energy calibration        &   2\%       &   1\%        &   2\%       &  3\%  \\
  \hline
  analysis stability            &   9\%       &  12\%        &   3\%       & 13\%        \\
  \hline
  p.d.f.s choice                &  10\%       &  15\%        &   1\%       &  2\%        \\
  \hline
  \end{tabular}
  \emph{\caption{Absolute value of the maximum relative changes on the 95\% 
  confidence level expected limits for each FCNC top decay branching ratio 
  evaluated with the likelihood-based analyses. The reference values were
  presented in Table~\ref{tab:lim95} ($L=10$~fb$^{-1}$).}
  \label{tab:syst}}
\end{center}
\end{table}

\section{Conclusions \label{conclusions}}

The sensitivity of the ATLAS experiment to the FCNC $t\to qX$
($X=Z,\gamma,g$) decays of the top quark was estimated. Different types of
analysis (cut-based and likelihood-based) were used to obtain the FCNC
branching ratio sensitivities (assuming a $5\sigma$ signal significance for
discovery) or the 95\% CL limits on the FCNC branching ratios (in the absence
of signal). The leptonic mode of the $t\to Z q$ channel was studied with
both type of analysis which give complementary results: the best limit on the
$BR$ assuming a signal discovery with a $5\sigma$ significance is obtained
with the cut-based analysis, while the 95\% CL limit obtained with the
likelihood-based analysis using the MFL method (which takes into account the
shape of the discriminant variables) is better. The impact of systematic
errors on the final results was also studied. The expected branching ratio
sensitivities obtained by the different analysis and the previous
ones~\cite{ATLAS,dodd,cakir} have the same order of magnitude, in the range
from $10^{-3}$ to $10^{-5}$ (for $L=10$~fb$^{-1}$). Even if the SM predicts a
much lower branching ratio for the FCNC decays of the top quark, the expected
branching ratios obtained in these analysis are several orders of magnitude
better then present experimental limits.

\begin{figure}
  \begin{center}
    \includegraphics[width=.895\textwidth]{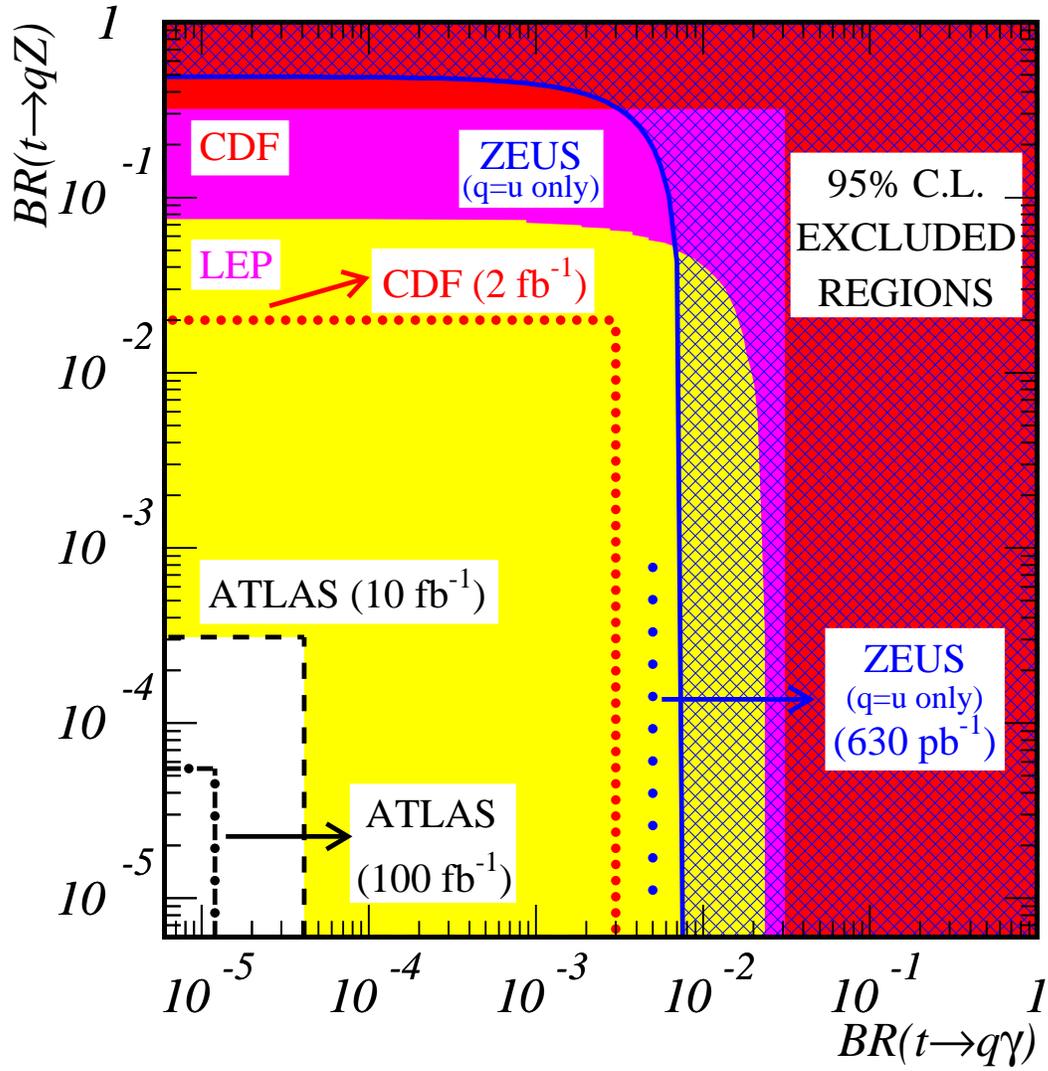}
  \end{center}

  \vspace*{-3em}
  \emph{\caption{The present 95\% CL limits on the $BR(t\to q\gamma)$
  vs. $BR(t \to qZ)$ plane are shown. The expected
  sensitivity at the HERA ($L=630$~pb$^{-1}$), Tevatron (run II) and LHC
  is also represented.}
  \label{fig:brplane}}
\end{figure}

The present 95\% CL limits and the expected sensitivity at the HERA (ZEUS,
$L=630$~pb$^{-1}$), Tevatron (CDF, run II~\cite{run2}) and LHC (ATLAS) for
$BR(t\to q\gamma,qZ)$ are summarised in Fig~\ref{fig:brplane}.


\section*{Acknowledgements}

We thank J.~A.~Aguilar-Saavedra, M.~Cobal, M.~David, P.~Ferreira,
D.~Froidevaux, C.~Marques, O.~Oliveira, E.~Richter-Was, R.~Santos and
S.~Slabospitsky for the very useful discussions. This work has been performed
within the ATLAS Collaboration, and we thank collaboration members for
helpful discussions. We have made use of the physics analysis framework and
tools which are the result of collaboration-wide efforts. This work was
supported by FCT -- Funda\c c\~ao para a Ci\^encia e a Tecnologia through the
grants SFRH/BD/13936/2003 and SFRH/BD/18762/2004.




\pagebreak


\clearpage

\end{document}